\documentclass[12pt]{article}
\usepackage[dvips]{graphicx}
\usepackage{amssymb}
\usepackage{amsmath}
\usepackage{epsfig}
\usepackage{cite}
\usepackage[pdfencoding=auto,psdextra]{hyperref}
\usepackage{bbold}
\usepackage{multirow}
\usepackage{subfig}

\numberwithin{equation}{section}
\numberwithin{table}{section}

\def\beq{\begin{equation}}
\def\eeq{\end{equation}}
\def\be{\begin{equation}}
\def\ee{\end{equation}}
\def\bea{\begin{eqnarray}}
\def\eea{\end{eqnarray}}

\def\Im{{\rm Im\,}}

\DeclareRobustCommand{\SkipTocEntry}[4]{}
\RequirePackage{color}
\begin{document}

\begin{titlepage}
\begin{center}
\rightline{\small }

%\begin{flushright} 
%\end{flushright}

\vskip 4cm

{\Large \bf Effective Theory of Warped Compactifications\\[.7em] and the Implications for KKLT
}
\vskip 1.2cm

Severin L\"ust, Lisa Randall
\vskip 0.5cm

{\small\it  Jefferson Physical Laboratory, \\
Harvard University, \\
17 Oxford St., \\
Cambridge, MA, 02138, USA.}
\vskip 0.1cm

\vskip 0.8cm

{\tt }

\end{center}

\vskip 1cm

\begin{abstract}

We argue that effective actions for warped compactifications can be subtle, with large deviations in the effective potential from naive expectations owing to constraint equations from the higher-dimensional metric. We demonstrate this deviation in a careful computation of  the effective potential for the conifold deformation parameter of the Klebanov-Strassler solution.  
The uncorrected naive effective potential  for the conifold was previously  used to argue that the Klebanov-Strassler background would be destabilized by antibranes placed at the conifold infrared tip unless the flux was uncomfortably large. We show this result is too strong because the formerly neglected constraint equations eliminate the features of the potential that allowed for the instability in the de Sitter uplift of the KKLT scenario.

\end{abstract}

\noindent

\vspace{0.2cm}

\noindent

\vfill

\noindent
%\today
June 2022

\end{titlepage}

%%%%%%%%%%%%%%%%%%%%%%%%%%%%%%%%%%%%%%%%%%%%%%%%%%%%

%\tableofcontents

%%%%%%%%%%%%%%%%%%%%%%%%%%%%%%%%%%%%%%%%%%%%%%%%%%%%

\section{Introduction}

Following the observation of an accelerated expansion of the Universe, the question of whether de Sitter vacua can be consistently realized as string theory backgrounds became one of the most pressing problems of string phenomenology.
While string theory vacua with a negative cosmological constant are ubiquitous, constructing a background with positive cosmological constant is notoriously difficult.
This is partially due to the lack of control which could otherwise be offered by supersymmetry.
It has furthermore been argued \cite{Dine:1985he,Maldacena:2000mw} that it is impossible to realize de Sitter vacua without taking into account strong coupling effects or fundamentally ``stringy'' ingredients such as D-branes and orientifold planes.
This moves them out of the realm of a pure supergravity analysis.

One of the most promising  proposals for realizing a string theory de Sitter vacuum is the KKLT scenario \cite{Kachru:2003aw}.
In this construction, a small parameter is used to uplift the negative energy of an AdS background. Though not necessarily essential, perhaps the most straightforward  source of such a parameter  (and the one found in the literature) is  a large RS like \cite{Randall:1999vf,Randall:1999ee} hierarchy arising from embedding a warped throat geometry into a compact Calabi-Yau space
along the lines of \cite{Verlinde:1999fy,Giddings:2001yu}.
Inserting a so-called anti-brane, which stabilizes at the bottom of this throat, would break supersymmetry and lift the cosmological constant to a positive value.
Due to the large gravitational redshift from the warped geometry, the contribution to the energy from the anti-brane would be exponentially suppressed. Cancelling much of the negative energy associated with the volume stabilization
 not only allows for an almost arbitrarily small cosmological constant, but also ensures that the construction remains under control.

Over the last two decades various potential issues with this proposal have been debated.
These include\footnote{The following list is necessarily incomplete, for a more exhaustive account see for example \cite{Danielsson:2018ztv} and references therein. There are also some recent more general objections against the existence of (long-lived) de Sitter vacua in string theory \cite{Obied:2018sgi,Bedroya:2019snp}.} the consistency of the probe-approximation to the antibrane uplift and possible singularities in its backreaction \cite{Bena:2009xk,McGuirk:2009xx,Dymarsky:2011pm,Bena:2011hz,Bena:2011wh,Bena:2012bk,Gautason:2013zw,Blaback:2014tfa,Michel:2014lva,Cohen-Maldonado:2015ssa,Bena:2022cwb}, the viability of a ten-dimensional description of K\"ahler moduli stabilization by non-perturbative effects \cite{Moritz:2017xto,Kallosh:2018wme,Gautason:2018gln,Hamada:2019ack,Carta:2019rhx,Kachru:2019dvo,Bena:2019mte,Grana:2020hyu}, control issues related to the embedding of a warped throat geometry into a compact Calabi-Yau manifold \cite{Carta:2019rhx,Gao:2020xqh,Carta:2021lqg}, the validity of the related four-dimensional EFTs \cite{Sethi:2017phn,Kachru:2018aqn,Lust:2022lfc} and the description of supersymmetry breaking therein \cite{DallAgata:2022abm}, as well as constraints on flux compactifications from tadpole cancellation conditions \cite{Braun:2020jrx,Bena:2020xrh,Bena:2021wyr,Marchesano:2021gyv,Plauschinn:2021hkp,Lust:2021xds,Grimm:2021ckh,Grana:2022dfw}.

The above have been the subject of debate and many issues have been at least partially addressed. 
However, \cite{Bena:2018fqc,Blumenhagen:2019qcg,Randall:2019ent,Bena:2019sxm,Dudas:2019pls} argued based on a straightforward potential analysis that embedding an anti-brane in a warped throat geometry, such as the Klebanov-Strassler solution \cite{Klebanov:2000hb}, can lead to a runaway instability of the geometry if the anti-brane is too heavy compared with the fluxes that stabilize the geometry (see also \cite{Bento:2021nbb} for a discussion of the weakly warped regime). 
This instability would give rise to an obstruction to the KKLT scenario unless one can ensure that the stabilizing fluxes can be chosen sufficiently large--leading to challenges for realizing the scenario and the requisite hierarchy. This analysis was much more straightforward so it was difficult to see where it  would fail. This simple analysis  was not subject to the more subtle inconsistency questions previously mentioned so in this respect seemed a bigger concern.

There were however reasons to doubt this conclusion. The holographic interpretation of the runaway direction would imply that  adding an antibrane  would prevent supersymmetry breaking and also would obstruct the gaugino condensate that is dual to the deformed conifold interpretation. It is unclear how the supergravity result can be consistent with what one would expect from a field theoretical perspective.

However, the analysis was based on the potential for the conifold deformation parameter in the literature \cite{Douglas:2007tu,Douglas:2008jx} with the addition of the potential from the antibrane including the overall warping. A strong clue to the resolution of the apparent paradox is that the computation of the instability relies heavily on the features of the off-shell potential for a geometrical modulus of the deformed conifold geometry that underlies the Klebanov-Strassler solution. 
Since the instability arises only if the effect of the anti-brane is of comparable importance, this requires an understanding of the potential not only in the direct, infinitesimal vicinity of its minimum, but also for finite field displacements.

We could also understand the issue  in terms of the low-energy effective theory.  Even if the complex structure moduli are stabilized at high energy through fluxes, the warped throat leads to light Kaluza Klein modes in the infrared with mass comparable to that of the conifold deformation parameter, whose potential is not necessarily as stable and whose values can change once the system is disturbed. Therefore the low-energy effective theory would include many more light fields that have conventionally been considered.

The effective actions of warped string compactifications is an old problem and has been for example earlier addressed in \cite{DeWolfe:2002nn,Giddings:2005ff,Frey:2006wv,Douglas:2007tu,Koerber:2007xk,Shiu:2008ry,Douglas:2008jx,Frey:2008xw,Marchesano:2008rg,Martucci:2009sf,Douglas:2009zn,Underwood:2010pm,Marchesano:2010bs,Grimm:2012rg,Frey:2013bha,Martucci:2014ska,Grimm:2014efa,Grimm:2015mua,Martucci:2016pzt}.
As first observed in \cite{Giddings:2005ff} (see also \cite{Douglas:2008jx} for a Hamiltonian interpretation)--if the KK modes are not explicitly included--every compactification on a background where the four-dimensional part of the space-time metric is multiplied by a warp factor that depends on the coordinates of the internal geometry has to be accompanied by additional constraint equations.
These constraints result in non-trivial relations between the various modes of the higher-dimensional background, in particular between the warp factor and the internal metric.

In this paper we demonstrate how to use these constraints to compute effective potentials for warped compactifications.
Specifically, in view of the aforementioned instability, we explicitly solve them to find the potential for the radion/conifold deformation parameter of the compactified Klebanov-Strassler geometry.
Here fluxes on the six-dimensional deformed conifold generate a large warp factor making it a prototypical laboratory for such backgrounds.
The underlying conifold geometry has one complex-structure modulus, namely its deformation parameter.
When embedded in a compact geometry the fluxes of the Klebanov-Strassler geometry generate a potential for this modulus, rendering it massive.

In previous attempts to compute the potential for this field the relevant constraints have been either ignored completely \cite{Douglas:2007tu} or  solved  in the hard-wall approximation \cite{Douglas:2008jx}. 
In the latter case the warp factor of the Klebanov-Strassler solution was approximated by that of a constant AdS geometry and its smooth IR cap was mimicked by a hard cutoff.
Here we use the exact Klebanov-Strassler warp factor and solve the constraint equations numerically.
We find the potential changes significantly in the infrared region of the throat in which the antibrane stabilizes.
Moreover, we find that the corrected off-shell potential resolves the before-mentioned instability.

This result is not only relevant for the stability of the KKLT construction but also illustrates that any effective field theory arising from warped compactifications can behave very 
differently than one might naively assume--particularly in the infrared region where Kaluza-Klein modes can be light, independently of the masses of the corresponding zero modes.
There are non-trivial relations between the zero and Kaluza-Klein modes of the different constituents of a warped background, in particular of the compactification manifold and the warp factor.
These effects can alter the form of an effective potential significantly.

In preparation for the derivation of the full off-shell potential, we first discuss the geometry of the deformed conifold in the absence of fluxes for which the background is a direct product of four-dimensional Minkowski spacetime and the conifold geometry.
I this case there is no potential for the geometric moduli of the conifold and the constraint equations simplify considerably and can be solved exactly.
As expected, we find that the deformation of the metric related to its complex structure modulus is localized in the infrared, near the tip of the conifold, and falls off asymptotically towards the ultraviolet, which is essential to modeling the strongly warped part of a string theory compactification using only the non-compact Klebanov-Strassler geometry.
Otherwise, the resulting effective action would be sensitive to the concrete UV embedding and its computation would require knowledge of the full compact Calabi-Yau geometry.

We subsequently use this insight to repeat the analysis of the constraint equations for the full warped Klebanov-Strassler solution.
Here we rely on numerical methods as the problem does not have any obvious analytical solution.
We find that the resulting deformation is again localized in the infrared and vanishes in the asymptotic ultraviolet.
Contrary to the non-compact Klebanov-Strassler setup, in an actual compactification, the fluxes  are quantized so we keep them constant when analyzing the deformations of the metric, allowing us to compute an explicit potential for the conifold deformation parameter.
The resulting potential differs qualitatively from the one obtained in \cite{Douglas:2007tu,Douglas:2008jx} that was used in \cite{Bena:2018fqc,Blumenhagen:2019qcg,Randall:2019ent,Bena:2019sxm} 
as it only has one critical point at the supersymmetric minimum both  with and without the antibrane perturbation.

We note however that there is yet another constraint equation in addition to the two we explicitly apply, formulated in \cite{Giddings:2005ff,Douglas:2009zn}, that becomes relevant only at higher order in the distance from a minimum of the potential.  
We argue that with the  potential we computed, the  vacuum expectation value of the conifold modulus that is induced by the antibrane shifts by only a small amount so the higher order effect can be neglected. Nonetheless it remains an interesting problem to solve all three constraints (and allow for flux deformation as well) to get the complete off-shell potential.

This paper is organized as follows: In Section~\ref{sec:conifoldrunaways} we review the conifold instability that was presented in \cite{Bena:2018fqc,Blumenhagen:2019qcg,Randall:2019ent,Bena:2019sxm}. In Section~\ref{sec:warpeddef} we summarize the relevant constraints on deformations of warped compactifications.
In Section~\ref{sec:unwarped} we illustrate how to utilize these constraints to obtain a family of metrics on the deformed conifold, related by complex structure deformations.
In Section~\ref{sec:KSpotential} we extend this analysis to non-vanishing fluxes and compute a potential for the complex structure modulus of the Klebanov-Strassler solution that we use in Section~\ref{sec:antibrane} to revisit the conifold instability.
In Section~\ref{sec:offshell} we explain the subtlety in a naive effective theory analysis and explain why it is essential to incorporate Kaluza-Klein modes from a low energy perspective. We also comment on the validity of the potential far away from its minimum. 
We conclude with some comments in Section~\ref{sec:conclusions} and three appendices.

\section{Review: conifold runaways from \texorpdfstring{$\overline{D3}$}{anti-D3}-branes in KKLT}\label{sec:conifoldrunaways}

The KKLT construction for de Sitter vacua \cite{Kachru:2003aw} builds on a warped type IIB flux background \cite{Grana:2000jj,Giddings:2001yu} with a ten-dimensional metric of the form
\begin{equation}\label{eq:10dwarped}
ds^2_{10} = e^{2 A(y)} \eta_{\mu\nu} dx^\mu dx^\nu + e^{-2A(y)} \tilde g_{mn}(y) dy^m dy^n \,.
\end{equation}
The warp factor $A$ depends only on the coordinates $y^m$ of the internal Calabi-Yau metric $\tilde g_{mn}$ and thus preserves the maximal symmetries of the four-dimensional metric $\eta_{\mu\nu}$.

Besides the ten-dimensional metric $g_{MN}$ the bosonic degrees of freedom of IIB supergravity comprise the Kalb-Ramond field $B_{MN}$, the dilaton $\Phi$ as well as even-degree form-fields $C_0$, $C_{MN}$ and $C_{MNPQ}$.
The two scalars $\Phi$ and $C_0$ are conveniently combined into one complex field $\tau_\mathrm{IIB} = C_0 - i e^{-\Phi}$, called the axio-dilaton.\footnote{In the following we reserve $\tau$ as the notation for the radial coordinate of the deformed conifold geometry, and hence denote the IIB axio-dilaton by $\tau_\mathrm{IIB}$.}

The field strengths of the form-fields $B_{MN}$, $C_{MN}$ and $C_{MNPQ}$ are denoted by $H_3$, $F_3$ and $F_5$ and can take non-trivial background values which are the internal fluxes. 
The three-form fluxes $F_3$ and $H_3$ are chosen in such a way that they  stabilize all the complex-structure moduli of the Calabi-Yau geometry. Their non-zero values  generate a strongly warped region via their back-reaction. 
The setup thus contains a feature reminiscent of a higher-dimensional realization of the Randall-Sundrum (RS) scenario \cite{Randall:1999vf,Randall:1999ee}, in which an AdS-like warped throat with a large gravitational redshift is glued to a compact Calabi-Yau geometry.
While the latter also naturally represents the UV-brane of RS, the IR-brane corresponds to a smooth cap at the end of the warped throat geometry \cite{Randall:2019ent}. 
Such a smooth warped throat is usually modeled by the Klebanov-Strassler (KS) solution \cite{Klebanov:2000hb}, a warped IIB solution over the deformed conifold with a particularly rich holographic interpretation.

With  the K\"ahler moduli  stabilized by non-perturbative effects,\footnote{This step involves the construction of an intermediate four-dimensional AdS-background.
Concerns against the viability of this approach related to its AdS/CFT holographic interpretation have been recently expressed in \cite{Lust:2022lfc}.} the large gravitational redshift can be exploited to facilitate the uplift to a small, positive cosmological constant in a controlled fashion.
This involves placing an $\overline{D3}$-brane, i.e.~a $D3$-brane with a charge incompatible with the supersymmetry of the background, in the throat, which naturally stabilizes in the  IR region at the bottom of the throat.
The large tension of the $\overline{D3}$-brane gets re-scaled by the exponentially large hierarchy between the IR and the UV and thus contributes only  a small amount to the overall energy density.
Therefore, if all the involved parameters are chosen correctly, one ends up with a controlled de Sitter background with an essentially arbitrarily small cosmological constant.

An important assumption in this construction is that the flux-generated masses of all complex-structure moduli are sufficiently high  that they can be effectively ignored when describing the stabilization of K\"ahler moduli and the antibrane-uplift.

That this assumption is potentially questionable can be seen from the observation that there is a distinct complex-structure modulus, a ``radion'', which controls the length and thus the hierarchy of the warped throat and can be shown to be equivalent to the conifold deformation parameter.
The wavefunction of this modulus is localized in the IR region  of the geometry.
Therefore, its mass is also set by the same exponential  redshift that reduces the antibrane's tension, rendering them of comparable scale.
One therefore has to pay particular attention to the interplay between the stabilization of this modulus and the $\overline{D3}$ uplift.

This problem has been discussed explicitly in \cite{Bena:2018fqc,Blumenhagen:2019qcg,Randall:2019ent,Bena:2019sxm}  for the example of the KS geometry.
Here, the local geometry of the Calabi-Yau metric $ds^2_6$ in the strongly warped region is given by the deformed conifold.
This non-compact space is defined by its embedding into $\mathbb{C}^4$ according to
\begin{equation}
    \sum_{i=1}^4 z^i = S \,,
\end{equation}
where the complex deformation parameter $S$ takes precisely the role of the before-mentioned radion.%
\footnote{That $S$ is indeed a complex-structure modulus can for example be seen from the A-cycle period integral
\begin{equation}
    \int_A \Omega = S \,,
\end{equation}
with $\Omega$ the holomorphic three-form on the deformed conifold. We discuss the role of $S$ as a complex-structure deformation in more detail in Appendix~\ref{app:cs}.}
The corresponding warp factor was found in \cite{Klebanov:2000hb} and reads
\begin{equation}\label{eq:AKS}
    e^{-4A(\tau)} = 2^\frac23 \frac{ g_s \left(\alpha' M\right)^2 }{ \left|S_0\right|^\frac43 } I(\tau) \,.
\end{equation}
Here, $\tau$ is a particular choice for a  ``radial'' coordinate on the deformed conifold that follows the warped geometry from its IR at $\tau = 0$ to the UV at $\tau \rightarrow \infty$.
$S_0$ denotes the value at which $S$ is stabilized, i.e.
\begin{equation}
    \left<S\right> = S_0 \,,
\end{equation}
and $M$ is the amount of $F_3$ flux along the A-cycle.
Moreover, $I(\tau)$ is a numerical function given by an integral expression, whereas $g_s$ and $\alpha'$ are the ten-dimensional string coupling constant and tension, respectively.
In terms of the coordinate $\tau$ the transition to the compact, weakly warped Calabi-Yau in the UV is effectively modeled by a cut-off $\tau_\Lambda$, related to some energy scale $\Lambda_0$ via
\begin{equation}\label{eq:Lambda0}
    \Lambda_0^3 \sim \left|S_0\right| e^{\tau_\Lambda} \,.
\end{equation}

As explained above, we are eventually interested in the mass and hence the stabilization potential for $S$.
While the superpotential is in principle well-known \cite{Vafa:2000wi} and holographically dual to the Veneziano-Yankielowicz superpotential \cite{Veneziano:1982ah,Taylor:1982bp}, the computation of the K\"ahler potential is more subtle as it depends non-trivially on the warp factor.
A corresponding K\"ahler metric has been suggested in \cite{Douglas:2007tu,Douglas:2008jx}, resulting in the potential
\begin{equation}\label{eq:VKS}
    V_{KS} \sim \frac{ \left|S\right|^\frac43 }{ g_s \left(\alpha' M\right)^2 } \left|\frac{M}{2 \pi i} \log \frac{\Lambda_0^3}{S} + i \frac{K}{g_s}\right|^2 \,,
\end{equation}
with $K$ the amount of $H_3$ flux on the B-cycle.
As advertised above, the overall potential is scaled by the warp factor $e^{4A}$ evaluated in the IR at $\tau=0$.
Moreover, one can easily see that there is a minimum at
\begin{equation}\label{eq:S0}
    S_0 = \Lambda_0^3 \exp \left( -\frac{2 \pi K}{ g_s M}\right) \,,
\end{equation}
which can be shown to be supersymmetric and can be made very small by choosing $M$ and $K$ accordingly.
Finally, there is also a contribution of the anti-brane to the potential for $S$ given by its warped-down tension
\begin{equation}\label{eq:VD3old}
    V_{\overline{D3}} = 2 e^{4A} T_{D3} \sim \frac{ \left|S\right|^\frac43 }{ g_s \left(\alpha' M\right)^2 } \,.
\end{equation}
A direct computation shows then that the combined potential $V_{KS} + V_{\overline{D3}}$  has a minimum only if
\begin{equation}\label{eq:Mbound}
    \sqrt{g_s} M > M_\mathrm{min} \,,
\end{equation}
where after reinstating and collecting all previously suppressed order-one constants the critical value can be estimated as $M_\mathrm{min} \approx 6.8 \sqrt{N_{\overline{D3}}}$ for $N_{\overline{D3}}$ antibranes \cite{Bena:2018fqc}.
Otherwise, the potential develops a runaway instability towards $S \rightarrow 0$, i.e.~an infinitely long throat, invalidating the KKLT construction.

It is crucial to notice that this instability relies heavily on the power-law dependence on $|S|$ in the inverse K\"ahler metric in \eqref{eq:VKS} and in the antibrane potential \eqref{eq:VD3old}.
This dependence in turn relies on the assumption that the ``off-shell'' $S$-dependence of the warp factor $e^{4A}$ when deforming the background away from the supersymmetric solution is the same as the dependence of $e^{4A}$ on the ``on-shell'' vev $S_0$ as in \eqref{eq:AKS}. 
In the remainder of this paper we will discuss this assumption and whether it is justified.

 \section{Off-shell fluctuations of warped compactifications}\label{sec:warpeddef}

Here we consider how the warped geometry changes when we take $S$ off-shell, which we refer to as a deformation of the warped geometry. We have seen that the existence of an instability of the type discussed in the previous section depends  on how the warp factor $e^{4A}$ varies as a function of the complex structure field $S$.
This dependence determines both the behavior of the flux-induced potential for $S$ \eqref{eq:VKS} and that of  the antibrane, \eqref{eq:VD3old} thereby determining the stability of $S$.

To discuss this in a bit more generality, we will first consider a general one-parameter family of background metrics, labeled by $S$.
This family consists of a warp factor and a six-dimensional Calabi-Yau metric,
\begin{equation}\label{eq:Agansatz}
    A(y^m, S) \qquad \text{and} \qquad \tilde g_{mn}(y^m, S) \,,
\end{equation}
such that at $S = S_0$ both agree with the ones in the background solution \eqref{eq:10dwarped}.
Moreover, we assume that all other fields do not depend on $S$ and stay at their background values, i.e.
\begin{equation}\label{eq:constantflux}
    \partial_S C_2 = \partial_S B_2 = \partial_S \tau_\mathrm{IIB} = 0 \,.
\end{equation}
We will later comment on the validity of this assumption, which we expect to break down if we go sufficiently far from the minimum of the potential 
such that a perturbative analysis is no longer valid.
Given the warp factor and metric as $S$-dependent functions as in \eqref{eq:Agansatz} we could then determine the potential for $S$ by inserting them into the ten-dimensional action and integrating over the six-dimensional internal part.

In previous analyses it was assumed that the $S$-dependence of the warp factor is given by
\begin{equation}\label{eq:offshellA}
    e^{-4 A(\tau, A) } \propto \frac{I(\tau)}{\left|S\right|^{\frac43}} \,,
\end{equation}
which is obtained from the ``on-shell'' warp factor \eqref{eq:AKS} by substituting $S$ for its vacuum expectation value $S_0$.
This not only assumes a very specific scaling behavior of the warp factor with respect to $S$, but also that its ``off-shell'' dependence on the radial coordinate $\tau$ stays the same as in the Klebanov-Strassler solution, given by the function $I(\tau)$.
Using \eqref{eq:offshellA} for the warp factor would indeed yield a potential of the form \eqref{eq:VKS}.

Naively, one might assume that we could take any family of warp factors (and metrics) \eqref{eq:Agansatz} with arbitrary dependence on $S$ so long as it matches the background solution at $S=S_0$, a condition clearly satisfied by \eqref{eq:offshellA}.
However, it was argued in \cite{Giddings:2005ff,Shiu:2008ry,Douglas:2008jx}  that not every such family is allowed and that, although the authors of these papers did not explicitly apply them on the KS solution, the dependence of $A$ and $\tilde g_{mn}$ on $S$ is restricted by various constraints.
Specifically, one observes that promoting $S$ to a four-dimensional dynamical field
\begin{equation}
    S \rightarrow S(x^\mu) \,,
\end{equation}
renders also the internal part of the ten-dimensional metric \eqref{eq:10dwarped} $x^\mu$ dependent by virtue of \eqref{eq:Agansatz}.
This in turn creates various additional terms in the ten-dimensional Einstein equations.
In the full ten-dimensional analysis some of them--in particular the off-diagonal $G_{\mu m}$ components--will  be satisfied  
for arbitrary $S(x^\mu)$ only if they are imposed as constraints, relating and restricting the variations $\partial_S A$ and $\partial_S \tilde g_{mn}$ in a non-trivial way.

Another way of satisfying these constraints is by introducing 
additional, compensating off-diagonal terms in the ten-dimensional metric.
Specifically, the ansatz for the ten-dimensional metric in the presence of a fluctuating field $S(x^\mu)$ should be extended according to
\begin{multline}\label{eq:compensatedmetric}
ds^2_{10} = e^{2 A(y, S)} g_{\mu\nu} dx^\mu dx^\nu \\+ e^{-2A(y, S)} \tilde g_{mn}(y, S) \left[dy^m - K^m(y, S) \partial_\mu S dx^\mu\right] \left[dy^n - K^n(y,S) \partial_\nu S dx^\nu \right] \,,
\end{multline}
where the compensators $K^m$ as well as the $S$-dependence of the warp factor are determined by solving the previously mentioned constraint equations.
One readily sees that the off-diagonal terms in the metric can be eliminated by a suitable redefinition of the internal coordinates $y^m$.
Doing so is equivalent to working in a specific gauge but changes the $S$-dependence of all fields implicitly.
For example, the variations of the warp factor and the metric change according to
\begin{equation}\begin{aligned}
    \delta_S A &= \partial_S A + \partial_m A K^m \,, \\
    \delta_S \tilde g_{mn} &= \partial_S \tilde g_{\mu\nu} + \widetilde \nabla_m \widetilde K_n + \widetilde \nabla_n \widetilde K_m \,,
\end{aligned}\end{equation}
where $\widetilde K_m = \tilde g_{mn} K^m$ and $\widetilde \nabla$ denotes the covariant derivative of $\tilde g_{mn}$.
Similar relations hold for all other fields in the theory.
Note that we have adopted the notation that $\delta_S$ denotes the variation of an object with respect to $S$ including the effect of the compensator $K^m$.
In particular, $\delta_S$ takes the role of a covariant derivative with respect to $S$ that is invariant with respect to diffeomorphism on the internal space.

Let us now return to the constraint equations on $\delta_S A$ and $\delta_S \tilde g_{mn}$
that follow from Einstein equations.
Denoting the Einstein tensor by $G_{MN} = R_{MN} - \frac12 g_{MN} R$, the relevant components read \cite{Giddings:2005ff,Shiu:2008ry}
\begin{equation}\begin{aligned}\label{eq:deltaeinstein}
\delta_S G_{\mu\nu} &= \partial_\mu \partial_\nu S \left(4 \delta_S A - \frac12 g^{mn} \delta_S \tilde g_{mn} \right) + \mathcal{O}(S^2) + \eta_{\mu\nu} \left[ \dots \right] \,, \\
\delta_S G_{\mu m} &= \partial_\mu S \left[2 \partial_m \delta_S A -\frac12 \partial_m \delta_S \tilde g - 8 \partial_m A \delta_S A + \partial_m A \delta_S \tilde g - 2 \partial^p A \delta_S \tilde g_{mn} + \frac12 \nabla^p \delta_S  \tilde g_{mp}  \right] \,.
\end{aligned}\end{equation}
In a sensible four-dimensional effective field theory $S$ must be allowed to fluctuate arbitrarily which means that the terms multiplying $\partial_\mu S$ or $\partial_\mu \partial_\nu S$ lead to constraints.
From the linear term in the first equation and the observation that no compatible term linear in $\partial_\mu\partial_\nu S$ can come from the $T_{\mu\nu}$ components of the stress energy tensor
one finds that the first constraint reads
\begin{equation}\label{eq:warpedtraceless}
\delta_S A = \frac18 \delta_S \tilde g \,,
\end{equation}
where $\delta_S \tilde g = g^{mn} \delta_S \tilde g_{mn}$.
It hence couples the variation of the warp factor and the trace of the internal metric.
This constraint implies that the warped volume 
\begin{equation}
    V_w  = \int d^6y \sqrt{\tilde g_6} e^{-4A}
\end{equation}
is conserved. 
This can be seen by noting that 
\begin{equation}
   \delta_S \left( \sqrt{\tilde g_6} e^{-4A} \right) = \partial_S \left( \sqrt{\tilde g_6} e^{-4A} \right) + \partial_m \left(\sqrt{\tilde g_6} e^{-4A} \tilde K^m\right) \,,
\end{equation}
and therefore $\partial_S V_m = 0$ up to a boundary term that vanishes on a compact internal space. 

From the second equation in \eqref{eq:deltaeinstein} one obtains another constraint,
\begin{equation}\label{eq:warpedoffdiagonal}
    \tilde \nabla^n \Bigl[ \delta_S \tilde g_{nm} -  \tilde g_{mn}  \left(\delta_S \tilde g - 4 \delta_S A \right) \Bigr] - 2 \tilde g^{nk} \, \partial_n A \Bigl[ \delta_S \tilde g_{km} - \tilde g_{km} \left( \delta_S \tilde g - 8  \delta_S A \right) \Bigr]= \delta_S T_m \,,
\end{equation}
where $\delta_S T_m$ denotes a possible contribution from the $T_{\mu m}$ stress energy tensor originating from the deformations of the additional fields: in our context $\delta_S C_2$, $\delta_S B_2$ and $\delta_S \tau_\mathrm{IIB}$.
This means that if these fields are kept constant, as in \eqref{eq:constantflux}, one has $\delta_S T_m=0$.
One can use the first constraint \eqref{eq:warpedtraceless} to eliminate the variation of the warp factor $\delta A$ from \eqref{eq:warpedoffdiagonal} to obtain
\begin{equation}\label{eq:warpedharmonic}
\tilde \nabla^n \left( \delta_S \tilde g_{nm} - \frac12 \tilde g_{mn} \delta_S \tilde g \right) - 4 \tilde g^{nk} \, \partial_n A \, \delta_S \tilde g_{km} = \delta_S T_m \,.
\end{equation}

In summary, we see that every deformation of the background must necessarily satisfy two constraint equations, \eqref{eq:warpedtraceless} and \eqref{eq:warpedoffdiagonal} that relate $\delta_S g_{mn}$ and $\delta_S A$.

 \section{Calabi-Yau metrics on the deformed conifold}\label{sec:unwarped}

We eventually want to use the constraints summarized in the previous section to derive an effective potential for the modulus $S$ in the strongly warped Klebanov-Strassler geometry.
However, before doing so we will first discuss the easier case in the absence of fluxes in which $S$ is a flat direction.
This will serve as a warp-up exercise on how to use the constraints.

The situation we are concerned with in this section is the compactification on a six-dimensional manifold with metric $g_{mn}$,
\begin{equation}\label{eq:10dnonwarped}
    ds^2_{10} = \eta_{\mu\nu} dx^\mu d x^\nu + g_{mn} dy^m dy^n \,,
\end{equation}
and no background fluxes; in the IIB context this means $F_3 = H_3 = F_5 = 0$ and $\tau_\mathrm{IIB} = \mathrm{const}$.
The equations of motion imply that $g_{mn}$ is Ricci flat.
In combination with the requirement of (partially) preserved supersymmetry this means that the internal manifold is Calabi-Yau.

The deformations of a Calabi-Yau metric which give rise to massless four-dimensional moduli fields are in principle well known. 
Specifically, they are given by infinitesimal deformations $g_{mn} \rightarrow g_{mn} + \delta g_{mn}$ such that the background stays Ricci-flat,
\begin{equation}\label{eq:ricciflat}
R_{mn}(g + \delta g) = 0 \,.
\end{equation}
When A is zero the constraints \eqref{eq:warpedtraceless} and \eqref{eq:warpedharmonic} readily reduce to the familiar gauge-fixing conditions \cite{Candelas:1985en}
\begin{equation}\label{eq:harmonictraceless}
\nabla^m \delta g_{mn} = 0 \,,\qquad g^{mn}  \delta g_{mn} = 0 \,.
\end{equation}
often referred to as transverse (or harmonic) traceless gauge.

On Calabi-Yau manifolds the deformations $\delta g_{mn}$ come in two different classes, K\"ahler and complex structure deformations.
They correspond to the two possible different index structures $\delta g_{i \bar \jmath}$ and $\delta g_{i j}$ with respect to complex coordinates $z^i$, $i = 1, \dots, 3$.
In the following we will mostly work only with real metrics, ignoring the existence of a complex structure, but revisit the distinction between K\"ahler and complex structure deformations in Appendix~\ref{app:cs}.
 
In the remainder of this section we discuss the conditions \eqref{eq:harmonictraceless} 
for the specific example of the deformed conifold. 
Here a further subtlety arises because we are dealing with a non-compact geometry.
Therefore, 
this setup would not give rise to an effective four-dimensional graviational theory without a five-dimensional cutoff.
In the KKLT context this is implemented by the assumption that the metric of the deformed conifold approximates the metric of another compact Calabi-Yau geometry in a certain region. 
As described in Section~\ref{sec:conifoldrunaways} this is modeled by a sufficiently large effective ``UV-cutoff'' for a non-compact coordinate $\tau$.
Consistency hence requires to only look at deformations of the deformed conifold that vanish in the UV, i.e.
\begin{equation}
    \delta g_{mn} \rightarrow 0 \qquad \text{for} \qquad \tau \rightarrow \infty \,.
\end{equation}

 To find the deformation of the metric of the deformed conifold we start with the following $SU(2) \times SU(2) \times \mathbb{Z}_2$ invariant ansatz \cite{Papadopoulos:2000gj},
 \begin{equation}\label{eq:conifoldmetricansatz}
 ds^2_6 = e^{-6 p(\tau) - x(\tau)} \left[d \tau^2 + (g^5)^2\right] + e^{x(\tau) + y(\tau)} \left[(g^1)^2 + (g^2)^2\right] +  e^{x(\tau) - y(\tau)} \left[(g^3)^2 + (g^4)^2\right] \,,
 \end{equation}
 where $g^1, \dots, g^5$ are one-forms defined in Appendix~\ref{app:conifoldframes}.
 While they span  five compact directions of the geometry, the before-mentioned coordinate $\tau$ parametrizes an a priori non-compact direction.
 However, it is bounded in the IR  at small $\tau$ by the conifold deformation and when embedded into a compact Calabi-Yau along the lines of \cite{Giddings:2001yu} also in the UV at large $\tau$.
Notice that this ansatz fixes the freedom to redefine $\tau \rightarrow \tilde \tau(\tau)$ up to a constant shift $\tau \rightarrow \tau + c$.

Interestingly, the problem to find functions $p(\tau)$, $x(\tau)$ and $y(\tau)$ such that the above metric is Ricci flat can be reformulated in terms of a one-dimensional action  \cite{Papadopoulos:2000gj},
\begin{equation}
S = \int d^6x \sqrt{g} R \sim \int d \tau \mathcal{L} \,,
\end{equation}
where the Lagrangian
\begin{equation}\label{eq:1dlagrangian}
\mathcal{L} = -\frac12 G_{ab} \dot \phi^a \dot \phi^b - V(\phi)
\end{equation}
 can be expressed in terms of a superpotential
 \begin{equation}
 V(\phi) = \frac18 G^{ab} \frac{\partial W}{\partial \phi^a} \frac{\partial W}{\partial \phi^b} \,.
 \end{equation}
Here we introduced the combined notation $\phi^a = (p, x, y)$.
Therefore, ``supersymmetric'' solutions  satisfy the first order equations
\begin{equation}\label{eq:susyeqs}
\dot \phi^a - \frac{1}{2} G^{ab}  \frac{\partial W}{\partial \phi^b} =0 \,.
\end{equation}
Direct computation of $R$ shows that
\begin{equation}
G_{ab} \dot \phi^a \dot \phi^b = e^{2x} \left(-\frac12 \dot x^2  + \frac12 \dot y^2 + 3 \dot p \dot x \right)
\end{equation}
and
\begin{equation}
W(\phi^a) = e^{-6p} + e^{2x} \cosh y \,.
\end{equation}
The set of first order equations resulting from \eqref{eq:susyeqs} has the following general solution
\begin{equation}\begin{aligned}\label{eq:conifoldsol}
e^{6p} &= \frac{6\left(\tfrac12 \sinh(2 \tau + 4 c_2) - \tau - 2 c_3\right)^{1/3}}{c_1^2 \sinh^2 ( \tau + 2 c_2) } \,, \\
e^{x} &= \frac{c_1}{2} \left(\tfrac12 \sinh(2 \tau + 4 c_2) - \tau - 2 c_3\right)^{1/3} \,, \\
e^{y} &= \tanh\left(\frac{\tau}{2} + c_2 \right) \,.
\end{aligned}\end{equation}
As expected from three first order equations, there are three constants of integration.
Moreover, the geometry ends at $\tau = -2 c_2$. 
In a holographic context this point corresponds to the IR of the dual field theory while the opposite direction $\tau \rightarrow \infty$ corresponds to the UV.
To obtain a smooth geometry we must require that the first derivatives of the metric vanish at the endpoint of the geometry.
Taking the $\tau$-derivatives of the three equations in \eqref{eq:conifoldsol} at $\tau \rightarrow 0$ one finds that 
this is only the case if $c_2 = c_3$.\footnote{If we were to consider the back-reaction of an antibrane or anything else with nontrivial energy-momentum tensor sitting at the end of the conifold, this condition could change.}
With this identification $c_2$ corresponds only to a shift in the coordinate $\tau$ and is hence a trivial diffeomorphism. 
We can therefore choose to eliminate it by setting it equal to zero, 
\begin{equation}
c_2 = c_3 = 0,
\end{equation}
which leaves us with only one constant of integration.
We will elaborate more on possible diffeomorphisms of $\tau$ later.

After identifying $c_1 = S^{2/3}$ with the deformation parameter of the deformed conifold
we now arrive at the familiar metric
\begin{multline}\label{eq:defconifoldmetric}
d s^2 = \frac{S^{2/3}}{2}  K(\tau) \biggl[  \frac{1}{3 K^3(\tau)} \left[ d \tau^2 + (g^5)^2 \right] 
+ \cosh^2\left(\frac{\tau} 2 \right) \left[ (g^3)^2 + (g^4)^2 \right] \\
+ \sinh^2\left(\frac{\tau} 2 \right) \left[ (g^1)^2 + (g^2)^2 \right]  \biggr] \,,
\end{multline}
where
\begin{equation}
K(\tau) = \frac{(\sinh(2 \tau) - 2 \tau)^{1/3}}{2^{1/3} \sinh \tau} \,.
\end{equation}
As we explain in Appendix~\ref{app:cs} this metric defines a Calabi-Yau metric on the deformed conifold.
Therefore, the first order equations \eqref{eq:susyeqs} indeed imply that the corresponding ten-dimensional background \eqref{eq:10dnonwarped} is supersymmetric.

Let us now try to use this form of the metric to determine its moduli, i.e.~its continuous deformations such that it stays Ricci flat. From the above analysis it follows that all such deformations must come from a variation in one of the integration constants, possibly combined with a change in the coordinate $\tau$, i.e.~a diffeomorphism.
Moreover, the only integration constant that keeps the metric smooth and is not just a pure diffeomorphism is $c_1 = S^{2/3}$. 

We therefore proceed by determining the variation of the metric that corresponds to a change in $S$,
\begin{equation}\begin{aligned}
\partial_S g_{mn} = \frac{\partial g_{mn}}{\partial S}  = \frac{2}{3 S} g_{mn} \,, \\
\end{aligned}\end{equation}
Clearly, $\partial_S g_{mn}$ is just a rescaling of the metric and therefore satisfies $\nabla^\mu \partial_S g_{mn} = 0$ but violates the second condition in \eqref{eq:harmonictraceless},
\begin{equation}
g^{mn} \partial_S g_{mn} = \frac{4}{S} \,.
\end{equation}

To transform $\partial_S g_{mn}$ into a harmonic and traceless deformation we have to combine it with a suitable diffeomorphism.
This means we would like to find a compensating vector $\eta^m$ such that
\begin{equation}\label{eq:deltaS}
\delta_S g_{mn} \equiv  \partial_S g_{mn} + (\mathcal{L}_\eta g)_{mn} =  \partial_S g_{mn} + \nabla_m \eta_n + \nabla_n \eta_m
\end{equation}
satisfies the gauge fixing conditions \eqref{eq:harmonictraceless}.
This results in two equations on $\eta^m$,
\begin{equation}\begin{aligned}\label{eq:gaugefixing}
\nabla^m \eta_m &= - \frac{1}{2 S} \,, \\
\nabla^m \nabla_m \eta_n &= 0 \,,
\end{aligned}\end{equation}
where in the last step we used the Ricc flatness of the metric.
Assuming that the corresponding change in coordinates should involve only the radial coordinate $\tau$
we employ the simple ansatz\footnote{See Appendix~\ref{app:cs} for a discussion of more general compensators with a non-vanishing $\psi$-component.}
\begin{equation}\label{eq:etaansatz}
\eta^m = \bigl( \eta^\tau(\tau), 0, 0,0,0,0 \bigr ) \,.
\end{equation}
The most general solution of the second order equation in \eqref{eq:gaugefixing} with this ansatz reads
\begin{equation}
\eta^\tau(\tau) = h_1 \frac{\sinh(2 \tau) - 2 \tau + h_2 }{\sinh^2 \tau } \,,
\end{equation}
where $h_1$ and $h_2$ are two integration constants.
The first equation in \eqref{eq:gaugefixing} fixes $h_1 = -1 / (2 S)$ and regularity at $\tau = 0$ requires $h_2 = 0$.

The solution for $\eta^\tau(\tau)$ describes an infinitesimal coordinate transformation for $\tau$.
We would also like to find the corresponding finite coordinate change
\begin{equation}\label{eq:newtau}
\tau \rightarrow \mathcal{T} \left(\tau, S\right) \,.
\end{equation}
Replacing $\tau$ with $\mathcal{T}$ in the metric and varying with respect to $S$ gives exactly the term $\mathcal{L}_{\eta} g_{\mu\nu}$ in \eqref{eq:deltaS} if
\begin{equation}
\frac{\partial \mathcal{T}(\tau, S) }{\partial S} = \eta^\tau \left[ \mathcal{T}(\tau, S) \right] \,.
\end{equation}
After a simple change of variables the right-hand side of this differential equation does not depend explicitly on $S$ anymore,
\begin{equation}
\frac{\partial \mathcal{T}}{\partial c} =  \frac{\sinh(2 \mathcal{T}) - 2 \mathcal{T} }{2 \sinh^2 \mathcal{T} } \,,\qquad c = -  \log S \,.
\end{equation}
and its solution is given by
\begin{equation}\label{eq:finitetauchange}
\mathcal{T}(\tau, S) = F^{-1} \left[F(\tau) -  \log \frac{S}{S_0} \right] \,,
\end{equation}
where $F^{-1}$ is the inverse function of
\begin{equation}
F(t) = \frac 12 \log \left[\sinh(2 t) - 2t \right] \,.
\end{equation}
Here the constant of integration is chosen such that $\mathcal{T}(\tau, S_0) = \tau$.

\begin{figure}[htb]
\centering\includegraphics[width=0.65\textwidth]{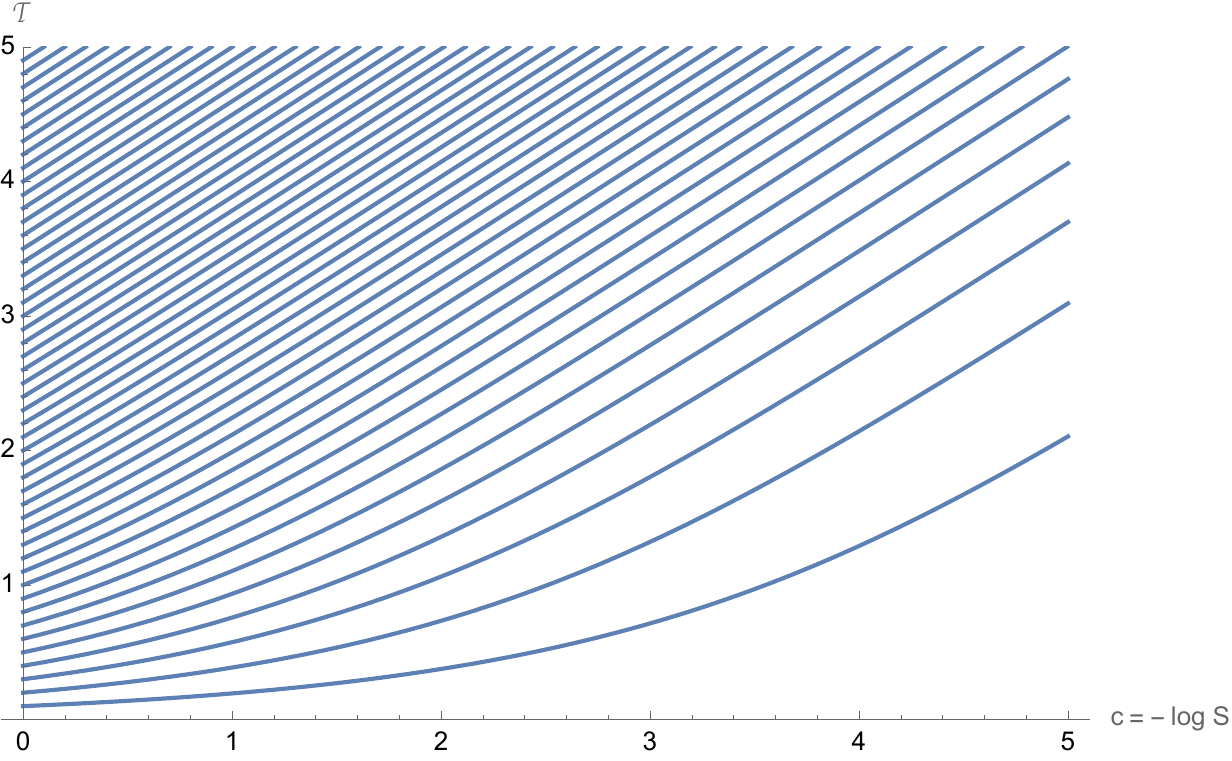}
\caption{The radial coordinate $\mathcal{T}$ as a function of $\log S$ (for $S_0=1$), where $S$ is the conifold deformation parameter.}
\label{fig:tauchange}
\end{figure}

We plot the evolution of the radial coordinate $\mathcal{T}(\tau, S)$ under a change of $S$ in Figure~\ref{fig:tauchange}.
To understand it further we can expand it in the IR and the asymptotic UV.
In the regime of small $\tau$ (corresponding to the IR) we can approximate $\mathcal{T}(\tau, S)$ by
\begin{equation}
\mathcal{T}(\tau, S) = \tau \left(\frac{S_0}{S}\right)^{2/3} + \mathcal{O}(\tau^3) \,,
\end{equation}
i.e.~it is given by a simple rescaling of $\tau$.
This yields the following expansion of the metric around $\tau = 0$,
\begin{multline}
    ds_6^2 = \frac{S^{2/3}}{2^{2/3} \times 3^{1/3}} \Bigl[ d \tau^2 + 2 (g^3)^2 + 2 (g^4)^5 + (g^5)^2 \Bigr] \\
    + \frac{\tau^2 }{5\times 2^{2/3} \times 3^{1/3} \, S^{2/3}} \left[ d \tau^2 + \frac52 (g^1)^2 + \frac52 (g^2)^5 + \frac32 (g^3)^2 + \frac32 (g^4)^5 + (g^5)^2 \right] + \mathcal{O}(\tau^4) \,.
\end{multline}
Therefore, at the tip of the conifold the two-sphere spanned by $g^1$ and $g^2$ shrinks to zero-size while the three-sphere spanned by $g^3$, $g^4$ and $g^5$ stays finite with a radius proportional to $S^{1/3}$.
We also see the dramatic effect of the gauge-compensator that gives rise to an inverse power of $S$ in the $\tau^2$-term.

For $\tau \rightarrow \infty$ (the UV), on the other hand, we find
\begin{equation}\label{eq:UVtau}
\mathcal{T}(\tau, S) \rightarrow \tau - \log \frac{S}{S_0} \,.
\end{equation}
To illustrate the significance of this behavior we notice that for $\tau \rightarrow \infty$ the metric asymptotes towards
\begin{equation}\label{eq:largetaumetric}
ds_6^2 \rightarrow \frac{3}{2^{5/3}} S^{2/3} e^{2 \mathcal{T} /3} \left[ \frac19 \left(d \mathcal{T}^2 + (g^5)^2 \right) + \frac16 \sum_{i = 1}^4 (g^i)^2  \right]  \,,
\end{equation}
the metric of the singular conifold.
With \eqref{eq:UVtau} the pre-factor satisfies 
\begin{equation}
S^{2/3} e^{2 \mathcal{T} /3} \rightarrow S_0^{2/3} e^{2  \tau /3} \,,
\end{equation}
so at $\tau \rightarrow \infty$ the metric is independent of $S$.

\begin{figure}[htb]
\centering
\subfloat[\label{fig:defmetric1}]{
\includegraphics[width=0.31\textwidth]{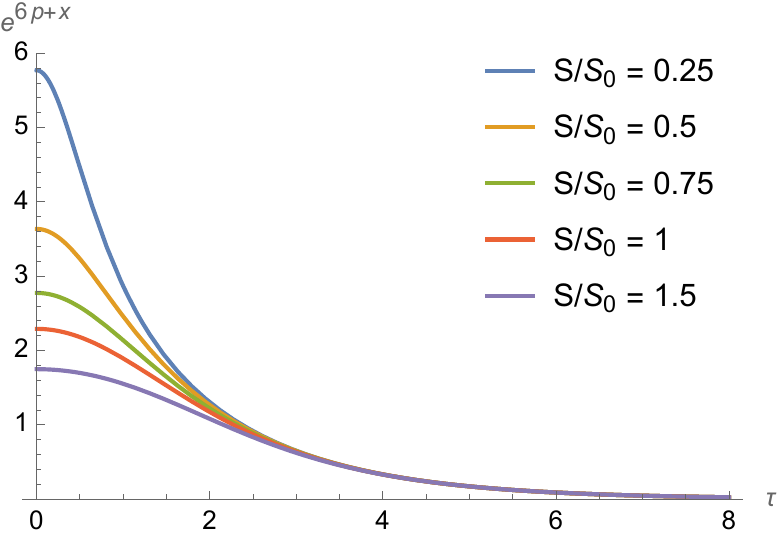}
}
%\hspace{0.03em}
\subfloat[\label{fig:defmetric2}]{
\includegraphics[width=0.31\textwidth]{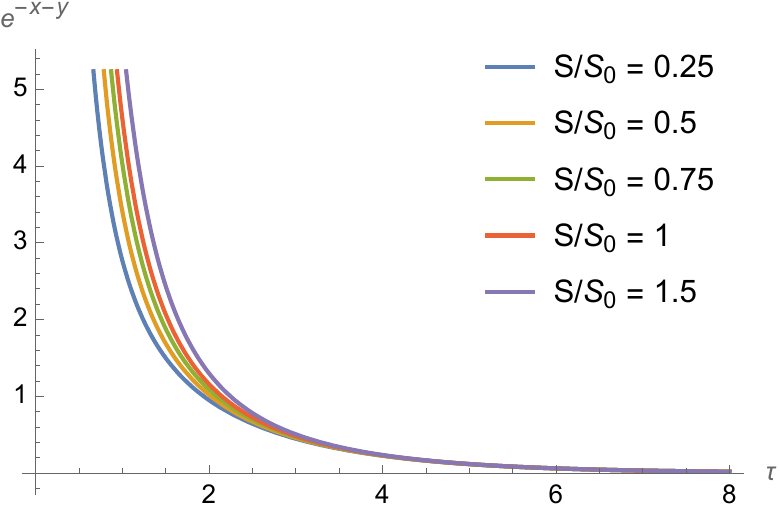}
}
%\hspace{0.03em}
\subfloat[\label{fig:defmetric3}]{
\includegraphics[width=0.31\textwidth]{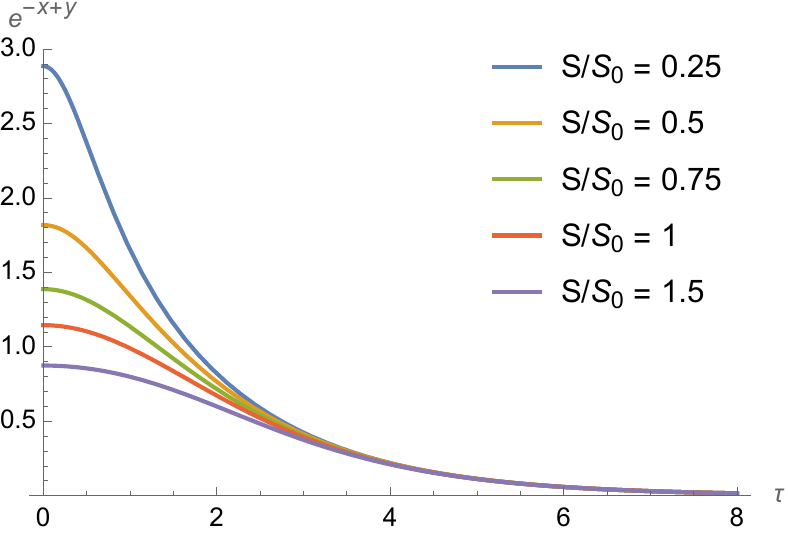}
}
\caption{The three independent components of the inverse metric on the deformed conifold for different values of $S/S_0$, taking into account the effect of the $S$-dependent diffeomorphism \eqref{eq:newtau} and \eqref{eq:finitetauchange}. (a) $g^{\tau\tau} = g^{55} = e^{6p+x}$, (b) $g^{11} = g^{22} = e^{-x+y}$, (c) $g^{33} = g^{44} = e^{-x-y}$.
}
\label{fig:defmetric}
\end{figure}

The full solution \eqref{eq:finitetauchange} interpolates between these two regimes 
and its effect on the three independent components of the metric under a change of $S$ is shown in Figure~\ref{fig:defmetric}.
It follows from the above discussion that the gauge-fixed variation of the metric is localized only in the IR region close to $\tau \approx 0$ and vanishes when approaching the UV at $\tau \rightarrow \infty$.
This is in stark contrast to the uncorrected variation with respect to $S$ which acts as a conformal prefactor of the metric and hence deforms it uniformly in the IR and the UV.
An essential part of the qualitative behavior of the deformation comes hence from the gauge transformation or diffeomorphism which is necessary to make the deformation harmonic and traceless.

As mentioned above, here we have treated the metric \eqref{eq:defconifoldmetric} on the deformed conifold as real and did not refer to a complex structure.
We have relegated this more technical discussion to Appendix~\ref{app:cs} where we show that $S$ is indeed a complex structure modulus and that $\delta_S g_{mn}$ arises from a harmonic $(2,1)$ form.
Specifically, the deformation we discussed here is the absolute value of the corresponding complex structure deformation.

We have seen (e.g.~in Figure~\ref{fig:defmetric}) that the metric deforms potentially significantly in the IR.
In the following section we will investigate how this affects the potential for the modulus field $S$ if we include non-trivial fluxes and hence a non-vanishing warp factor $A(\tau)$. 
Analogous to the discussion here the deformations of the metric and now also the warp factor will be determined by the constraint equations.

We anticipate that solving these equations will significantly modify the IR behavior of the metric and warp factor, and these will be relevant to the low-energy potential for the light fields with amplitude concentrated in the IR.

\section{Application to the Klebanov-Strassler solution}\label{sec:KSpotential}

We are now in a position to apply this formalism to the warped deformed conifold, i.e.~the Klebanov-Strassler solution.
Let us first collect some basic facts about the background we are going to deform. 
The ten-dimensional metric is of the form \eqref{eq:10dwarped} with the internal metric $\tilde g_{mn}$ given by the Ricci flat metric on the deformed conifold \eqref{eq:defconifoldmetric}.
The warp factor $A(\tau)$ is given by (note that it is the inverse square root of this expression which appears in \eqref{eq:10dwarped} in front of the four-dimensional part of the metric)
\begin{equation}\label{eq:AKSagain}
    e^{-4A(\tau)} = 2^\frac23 \frac{ g_s \left(\alpha' M\right)^2 }{ \left|S_0\right|^\frac43 } I(\tau) \,,
\end{equation}
where all $\tau$-dependence comes from the function $I(\tau)$ that is given by the integral expression
\begin{equation}\label{eq:IKS}
    I(\tau) = \int^\infty_\tau dx \frac{x \coth x - 1}{\sinh^2 x} \left(\sinh 2x - 2x\right)^{1/3} \,.
\end{equation}
Moreover, there is the self-dual three-form flux
\begin{equation}\label{eq:g3}
G_3 = F_3 - \frac{i}{g_s} H_3 = \sqrt{6} M \alpha' \chi_S \,,
\end{equation}
where $\chi_S$ is the unique harmonic $(2,1)$ form on the deformed conifold, see \eqref{eq:chiS} in Appendix~\ref{app:cs}.

These fluxes source not only a non-trivial warp factor but also the self-dual five-form flux
$F_5 = \mathcal F_5 + \star \mathcal F_5$.
Four-dimensional Poincar\'e invariance demands that
\begin{equation}\label{eq:dC4}
    \mathcal F_5 = d C_4 = d \alpha \wedge \mathrm{vol}_4 \,,
\end{equation}
for $\alpha = \alpha(\tau)$ and the Bianchi identity enforces
\begin{equation}\label{eq:F5B2F3}
\star \mathcal F_5 =  B_2 \wedge F_3 \,.
\end{equation}
The supersymmetric equations of motion are solved for $\alpha = e^{4A}$ (see e.g. \cite{Giddings:2001yu}) and therefore
\begin{equation}\label{eq:Afluxes}
    \tilde \star_6\, d e^{-4A} = \star\, \left(d e^{4A} \wedge \mathrm{vol}_4\right)  = B_2 \wedge F_3 \,.
\end{equation}
Here we used that the transition from the Hodge-$\star$ operator of the full ten-dimensional metric to the six-dimensional $\tilde \star_6$ of the internal Calabi-Yau metric gives an additional factor of $- e^{-8A}$.
The relation \eqref{eq:Afluxes} can be used to determine the warp factor $e^{-4A}$ in \eqref{eq:AKS} as the integral of $ \tilde \star_6 (B_2 \wedge F_3)$. 
For the KS-solution the latter expression can be evaluated explicitly and leads to the integral expression \eqref{eq:IKS}.

The integer $M$ in \eqref{eq:g3}  determines the amount of $F_3$ flux over the A cycle of the deformed conifold,
\begin{equation}
M = \frac{1}{(2 \pi)^2 \alpha'} \int_A F_3 \,.
\end{equation}
On the other hand, the B cycle of the deformed conifold stretches along its non-compact direction.
Therefore we need a UV cutoff $\Lambda_0$, see \eqref{eq:Lambda0}, to obtain a finite $H_3$ flux $K$ over the B-cycle,
\begin{equation}
K = \frac{1}{(2 \pi)^2 \alpha'} \int_A H_3 = \frac{1}{(2 \pi)^2 \alpha'} \int_0^{\tau_0} d \tau \int_{S^2} H^3 \approx \frac{g_s M}{2 \pi} \log \left(\frac{ \Lambda_0^3}{S_0}\right) \,,
\end{equation}
where in the last step we assumed that $\Lambda_0^3 \gg S$.
Therefore, even though we are in principle able to choose $S$ freely, fixing $M$ and $K$ as well as the UV cutoff $\Lambda_0$ will create a potential for $S$ with a minimum at the value $S_0$ given in \eqref{eq:S0}.

To determine the potential for $S$ we need first to establish which parts of the ansatz to keep fixed and which to vary as we change $S$.
As already explained in Section~\ref{sec:warpeddef}, here we choose to keep the two-form fields and the axio-dilaton exactly constant,
\begin{equation}\label{eq:constantfluxes}
\delta_S C_2 = \delta_S B_2 = \delta_S \tau_\mathrm{IIB} = 0 \,, 
\end{equation}
but consider a one-parameter family of warp factors and Ricci-flat metrics on the deformed conifold,
\begin{equation}\label{eq:Sfamily}
    A = A(\tau, S) \,,\qquad \tilde g_{mn} = \tilde g_{mn} (\tau, S) \,.
\end{equation}
Importantly, we cannot choose the dependence of $F_5$ (or equivalently $\alpha$, see \eqref{eq:dC4}) on $S$ freely.
For constant $B_2$ and $F_3$ it is uniquely determined by the action of the Hodge-$\star$ operator in \eqref{eq:F5B2F3}.
Moreover, as discussed in Section~\ref{sec:warpeddef}, we should in principle also allow for an additional compensator $K_m$.
Such a compensator, however, would also enter the conditions \eqref{eq:constantfluxes}.
For example the variation of $C_2$ reads
\begin{equation}
    \delta_S C_2 = \partial_S C_2 + \mathcal{L}_K C_2 \,,
\end{equation}
where $\mathcal{L}$ denotes the Lie-derivative.
Therefore, \eqref{eq:constantfluxes} implies that $\partial_S C_2 = - \mathcal{L}_K C_2$ and hence we can perform a coordinate redefinition after which $\partial_S C_2 =0$ and
\begin{equation}
    K_m = 0 \,.
\end{equation}
The same holds argument holds for $B_2$ and $\tau_\mathrm{IIB}$.
Such a change of coordinates of course also affect the--so far undetermined--warp factor $A(\tau, S)$ and metric $ \tilde g_{mn}$ but can be absorbed into their definition. 
Therefore \eqref{eq:Sfamily} represents the most general ansatz that is compatible with \eqref{eq:constantfluxes}.

We have seen in the previous section that a family of Ricci flat Calabi-Yau metrics labeled by the complex structure parameter $S$ on the warped deformed conifold takes the form \eqref{eq:defconifoldmetric} up to an $S$-dependent diffeomorphism of the radial coordinate $\tau$.
A suitable ansatz for $\tilde g_{mn}$ hence reads\footnote{For further comments on the generality of this ansatz see Appendix~\ref{app:hardwall}.}
\begin{multline}\label{eq:gmnofS}
\tilde g_{mn} dy^m dy^n = \frac{S^{2/3}}{2}  K \bigl(\mathcal{T}(\tau, S)\bigr) \Biggl[  \frac{1}{3 K\bigl(\mathcal{T}(\tau, S)\bigr)^3} \left[ \bigl(\partial_\tau \mathcal{T}\bigr)^2 d \tau^2 + (g^5)^2 \right] \\
+ \cosh^2\left(\frac{\mathcal{T}(\tau, S)} 2 \right) \left[ (g^3)^2 + (g^4)^2 \right] 
+ \sinh^2\left(\frac{\mathcal{T}(\tau, S)} 2 \right) \left[ (g^1)^2 + (g^2)^2 \right]  \Biggr] \,,
\end{multline}
where besides the conformal prefactor of $S^{2/3}$ the yet to be determined function $\mathcal{T}(\tau, S)$ encodes all the non-trivial $S$ dependence of $\tilde g_{mn}$. 

Our family of configurations \eqref{eq:Sfamily} is thus reduced to two functions $A(\tau, S)$ and $\mathcal{T}(\tau, S)$.
The former two are not independent but related by the constraint \eqref{eq:warpedoffdiagonal}.
To utilize this constraint we notice that it can be rewritten as \cite{Giddings:2005ff}
\begin{equation}\label{eq:geometricconstraint2}
\left[ \tilde \star_6 \delta_S \left(\tilde \star_6 d e^{-4 A} \right) \right]_m + e^{4A} \tilde \nabla^n \left( \delta_S \tilde g_{nm} -  \tilde g_{mn} \delta_S \tilde g \right)= \delta_S T_m  \,.
\end{equation}
Moreover, the ansatz \eqref{eq:gmnofS} for the metric satisfies
\begin{equation}
    \tilde \nabla^m \delta_S \tilde g_{mn} - \tilde \nabla_n \delta_S \tilde  g = 0 \,,
\end{equation}
for any $\mathcal{T}(\tau, S)$
and the condition \eqref{eq:constantfluxes} implies $\delta_S T_m = 0$.
Therefore, \eqref{eq:geometricconstraint2} reduces to
\begin{equation}
    \delta_S \left(\tilde \star_6 d e^{-4A}\right) = 0 \,.
\end{equation} 
We can use this relation to conclude that $\tilde \star_6 d e^{-4A}$ remains constant along the flow of $S$.
Since the initial value of $d e^{-4A}$ at $S=S_0$ is given by \eqref{eq:Afluxes}, we find that
\begin{equation}\label{eq:integratedconstraint}
    d e^{-4A(S)} = \tilde \star_6  \left(B_2 \wedge F_3\right) \,,
\end{equation}
and
\begin{equation}\label{eq:alphaoffshell}
    \alpha(\tau, S) = e^{4A(\tau, S)} \,,
\end{equation}
for any value of $S$.
We can understand \eqref{eq:integratedconstraint} as an integrated version of the constraint \eqref{eq:geometricconstraint2}.
Moreover, as also observed in \cite{Shiu:2008ry}, the relations \eqref{eq:integratedconstraint} and \eqref{eq:alphaoffshell} reduce the potential to the familiar expression
\begin{equation}\label{eq:GGpotential}
V_\mathrm{flux} = \frac{1}{4 \kappa_{10}^2 } \int d^6y \frac{e^{4A}}{\Im \tau_\mathrm{IIB}} \Bigl[G_3 \wedge \tilde \star_6 \overline{G}_3 + i G_3 \wedge  \overline{G}_3 \Bigr] \,.
\end{equation}
The potential in \eqref{eq:GGpotential} depends on $S$ via the warp-factor $e^{4A}$ and through the $\tilde\star_6$-operator via the internal metric $\tilde g_{mn}$. 

For $B_2$ and $F_3$ given by the Klebanov-Strassler solution \eqref{eq:g3} and \eqref{eq:chiS} we can give an explicit expression for the integrated constraint \eqref{eq:integratedconstraint},
\begin{equation}\label{eq:harmonicconstraintsol}
    \partial_\tau e^{-4A(\tau, S)} = 2^\frac23 \frac{ g_s \left(\alpha' M\right)^2 }{ \left|S\right|^\frac43 } \frac{\tau \coth \tau - 1}{\sinh^2 \tau} \frac{\sinh 2 \tau - 2\tau}{\left(\sinh 2 \mathcal{T} - 2\mathcal{T}\right)^{2/3}} \partial_\tau \mathcal{T} \,.
\end{equation}
At $S=S_0$, where $\mathcal{T}(\tau, S_0) = \tau$, this consistently reduces to the Klebanov-Strassler warp factor \eqref{eq:IKS}.
Moreover, we have seen in Section~\ref{sec:unwarped} that the metric of the deformed conifold stays asymptotically constant in the UV  if
\begin{equation}\label{eq:TUV}
    \mathcal{T}(\tau, S) \rightarrow \tau - \log \frac{S}{S_0} \qquad \text{for} \qquad \tau \rightarrow \infty \,.
\end{equation}
Inserting this into \eqref{eq:harmonicconstraintsol} shows that
\begin{equation}
    \partial_\tau e^{-4A(\tau, S)} \rightarrow 2 \frac{ g_s \left(\alpha' M\right)^2 }{\left|S_0\right|^\frac43 } e^{-4 \tau/3} \left(\tau-1\right) \,,
\end{equation}
for $\tau \rightarrow \infty$.
The right hand side does not depend on $S$ but only on $S_0$ and so we find that also  $\delta_S A \rightarrow 0$ in the UV.
In other words, the asymptotic behavior \eqref{eq:TUV} ensures that both $\tilde g_{mn}$ and $A$ fluctuate with $S$ only in the IR and that our fluctuated KS geometry can still be embedded into a compact Calabi-Yau geometry along the lines of \cite{Giddings:2001yu}. 

By inserting the relation \eqref{eq:harmonicconstraintsol} back into the IIB action we can explicitly compute the potential \eqref{eq:GGpotential} and find the (relatively bulky) expression
\begin{equation}\begin{gathered}\label{eq:explicitpotential}
    V_\mathrm{flux} = T_3 \frac{g_s M^2}{32 \pi} \int d \tau \frac{e^{4A}}{\partial_\tau \mathcal{T}} \Biggl\{ 8 \frac{\left(\tau \coth\tau - 1\right)^2}{\sinh^2 \tau} + \coth^2 \left(\mathcal{T}/2\right) \frac{\sinh^2( \tau/2)}{\cosh^6(\tau/2)} \left(\sinh \tau + \tau\right)^2 \\
    + \tanh^2 \left(\mathcal{T}/2\right) \frac{\cosh^2( \tau/2)}{\sinh^6(\tau/2)} \left(\sinh \tau + \tau\right)^2
    + 16 \left[1 + \frac{3 + 2 \tau - 6 t \coth \tau + 3 \tau^2 \operatorname{csch}^2 \tau}{\sinh^2 \tau }\right] \partial_\tau \mathcal{T} \\
    + 8 \left[1 + 2 \operatorname{csch}^2 \mathcal{T} - 4 \frac{\cosh \mathcal{T}}{\sinh^2 \mathcal{T}} \operatorname{csch} \tau  +  \frac{ (\tau \coth \tau -1)^2 + \tau^2 (1 + 2 \operatorname{csch}^2 \mathcal{T}) }{\sinh^2\tau} \right] \left(\partial_\tau \mathcal{T}\right)^2
    \Biggr\} \,.
\end{gathered}\end{equation}
Here we used again the expressions \eqref{eq:g3} and \eqref{eq:chiS} for the three-form flux $G_3$ in the KS solution.
Also, when integrating over the remaining five compact dimensions we had to account for the correct numerical factor \eqref{eq:5dvolume}.
As a consistency check it can be verified that the integrand in \eqref{eq:explicitpotential} vanishes identically if $\mathcal{T} = \tau$ and hence the minimum at $S=S_0$ gives rise to Minkowski space.

\begin{figure}[htb]
\centering
\subfloat[\label{fig:warpfactors2}]{
\includegraphics[width=0.45\textwidth]{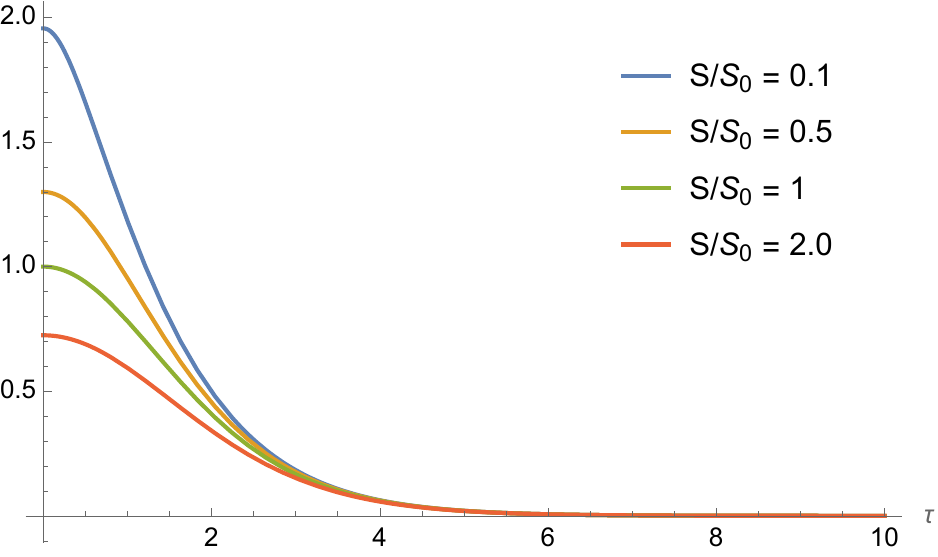}
}
\hspace{0.4em}
\subfloat[\label{fig:integrandprofile}]{
\includegraphics[width=0.45\textwidth]{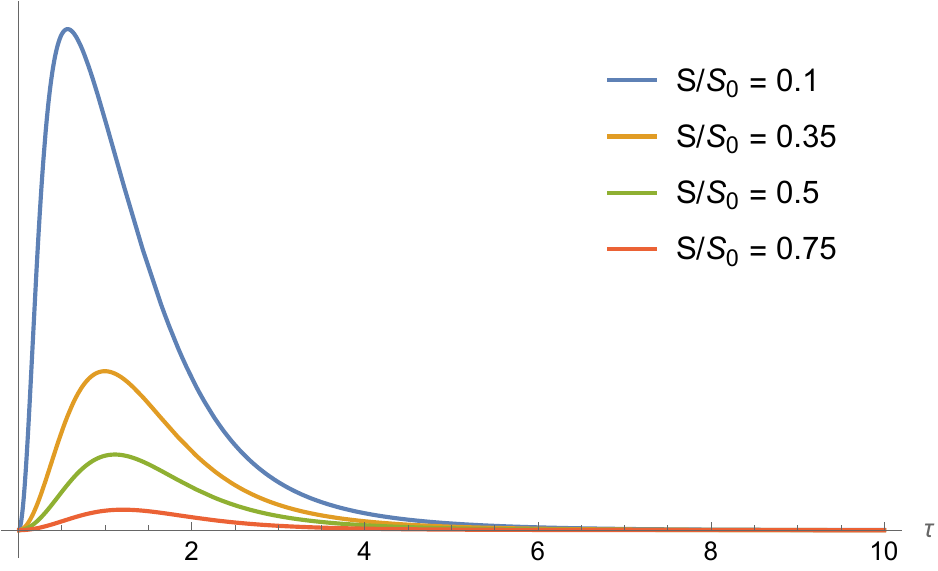}
}
\caption{(a) The inverse warp factor $e^{-4A}$ as a function of the radial coordinate $\tau$ for different values of $S/S_0$.
(b) The integrand of the potential \eqref{eq:explicitpotential} for different values of $S/S_0$.
In both figures the IR-tip of the throat geometry is located at $\tau = 0$ and $\tau \rightarrow \infty$ corresponds to its UV.
}
\label{fig:warpfactors}
\end{figure}

To compute the four-dimensional potential $V_\mathrm{flux}(S)$ as a function of $S$ from \eqref{eq:explicitpotential}, it remains to determine $\mathcal{T}(\tau, S)$ and $A(\tau, S)$.
This is done by exploiting also the first constraint
\eqref{eq:warpedtraceless}.
In our case it yields the relation
\begin{equation}
 \frac2S + \delta_S \Bigl[ \log \left( \sinh^2 \mathcal{T} +  \partial_\tau \mathcal{T} \right)- 4  A\Bigr]   = 0 \,,
\end{equation}
which can be directly integrated to obtain
\begin{equation}\label{eq:volumeconstraintsol}
    \frac{e^{-4A(S)}}{e^{-4A(S_0)}} \left(\frac{S \sinh \mathcal{T}}{S_0 \sinh{\tau}}\right)^2  \partial_\tau \mathcal{T} = 1 \,.
\end{equation}
In combination with the other constraint \eqref{eq:harmonicconstraintsol} this relation yields a system of differential equations for $A$ and $\mathcal{T}$ that has to be solved numerically.
For example, we can use \eqref{eq:volumeconstraintsol} to eliminate $e^{-4A}$ from \eqref{eq:harmonicconstraintsol} to obtain an ordinary differential equation of second degree for $\mathcal{T}$ at every value of $S$.
When solving this differential equation initial conditions have to be imposed such that $\mathcal{T}(\tau, S)=0$ at $\tau = 0$ and such that it goes asymptotically towards \eqref{eq:TUV} at large $\tau$. 
Once $\mathcal{T}(\tau, S)$ is determined one can solve \eqref{eq:volumeconstraintsol} for $A(\tau, S)$ and compute the potential by numerically integrating \eqref{eq:explicitpotential}.

\begin{figure}[htb]
\centering
\includegraphics[width=0.75\textwidth]{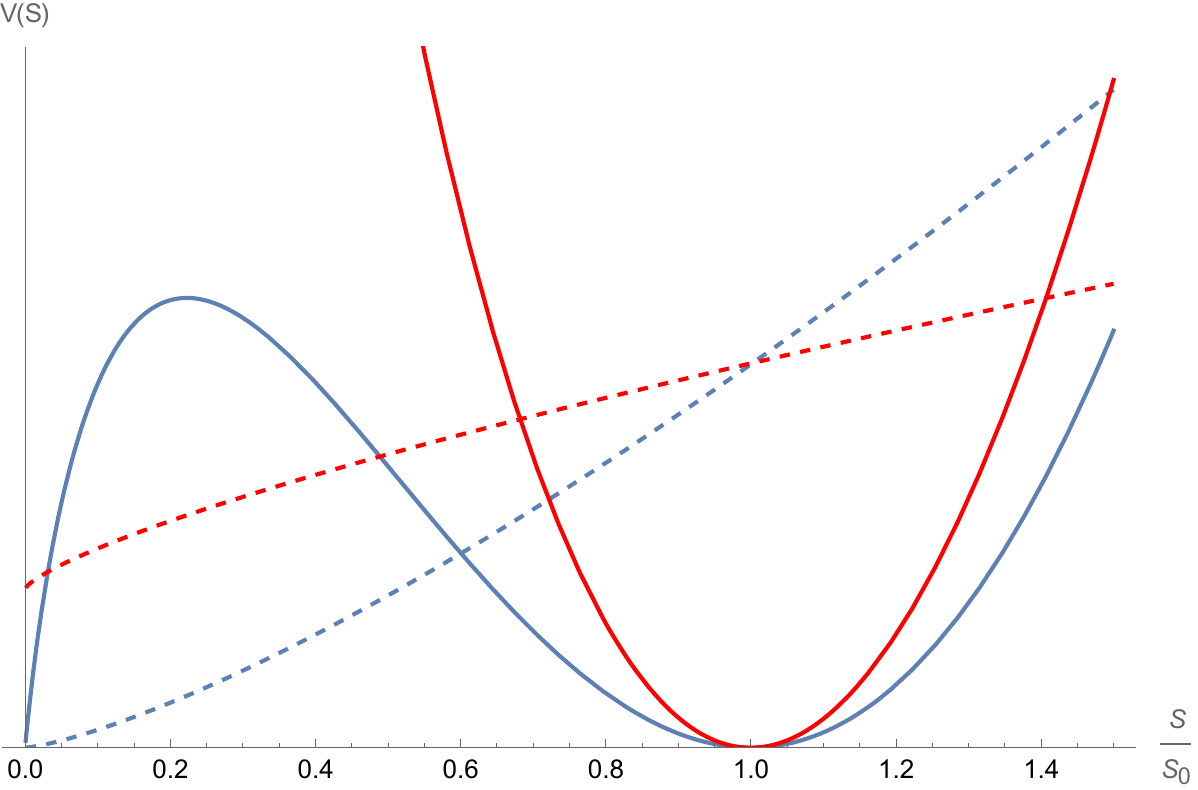}
\caption{This plot compares previous results for the potential for the conifold modulus $S$ with ours. The solid lines are the flux potential \eqref{eq:GGpotential} at $g_s^{1/2} M =10$  and the dashed lines the contribution from the antibrane.
The potential computed by \cite{Douglas:2008jx} is in blue and our potential in red.
Their superposition is illustrated in Figure~\ref{fig:potentialsuperposition}.}
\label{fig:potentialcomparison}
\end{figure}

The resulting profile of the warp factor is shown in Figure~\ref{fig:warpfactors2}.
We  see that  not only does the warp factor change only in the IR where $\tau$ is small,  but also that in this region the dependence of $e^{-4A}$ on $S$ is much weaker than the naive expectation $e^{-4A} \sim S^{-4/3}$. 
Around the minimum for $S \approx S_0$ and at the IR tip of the conifold we can approximate it by
\begin{equation}
e^{-4A(\tau = 0, S)} \propto \left(\frac{S_0}{S}\right)^{0.44} \,.
\end{equation}
Moreover, for $S\rightarrow 0$ the warp factor appears to asymptote to a constant value and does not diverge.
Therefore, we expect the potential $V(S)$ not to have any additional critical points and not to go back to zero for $S\rightarrow 0$. 
We confirm this expectation by  inserting the result for $A(\tau, S)$ into \eqref{eq:explicitpotential} and performing the $d \tau$ integral numerically.
We illustrate the profile of the integrand for different values of $S/S_0$ in Figure~\ref{fig:integrandprofile}.
The resulting potential can be found in Figure~\ref{fig:potentialcomparison}.

\section{Effect on the conifold instability}\label{sec:antibrane}

Let us finally discuss the effect of the corrected potential for $S$ on the conifold instability found in \cite{Bena:2018fqc,Blumenhagen:2019qcg,Randall:2019ent,Bena:2019sxm}.
As briefly summarized in Section~\ref{sec:conifoldrunaways}, adding an anti-brane in the IR of the Klebanov-Strassler geometry gives an additional contribution to the potential for $S$.
This potential can be computed from the antibrane action 
\begin{equation}
    S_{\overline{D3}} = S_{DBI} + S_{CS} = - T_3 \int d^4x \sqrt{-g_4} - T_3 \int C_4 \,,
\end{equation}
with $T_3$ the D3-brane tension.
In our case, where \eqref{eq:alphaoffshell} holds, the two terms give exactly the same contribution.
By evaluating this action in the IR at $\tau = 0$ where the antibrane sits we find 
\begin{equation}\label{eq:VD3}
    V_{\overline{D3}}(S) = 2 N_{\overline{D3}} T_3 e^{4A(0, S)} \,,
\end{equation}
with $N_{\overline{D3}}$ the number of antibranes. 
This means that the functional dependence of $V_{\overline{D3}}$ on $S$ is just given by the behavior of the warp factor at the position of the anti-brane, namely the IR end of the Klebanov-Strassler throat.

The total potential for $S$ consists then of two contributions, the flux potential computed in the previous section and the antibrane potential,
\begin{equation}\label{eq:combinedpotential}
    V(S) = V_\mathrm{flux}(S) + V_{\overline{D3}}(S) \,,
\end{equation}
with $V_\mathrm{flux}(S)$ given in \eqref{eq:explicitpotential}.
While $V_\mathrm{flux}(S)$ has a minimum at $S=S_0$, the antibrane term $V_{\overline{D3}}$ is a monotonically increasing function of $S$ that is minimized at $S=0$.
Therefore, it will drive $S$ to smaller values and the minimum of the combined potential will be at $0 \leq S < S_0$.
The magnitude of this shift and whether it can destabilize the conifold all the way to the singular conifold at $S=0$ depends on two factors.
Firstly, the flux potential \eqref{eq:explicitpotential} is enhanced by a relative factor of $g_s M^2$ which is not present in the antibrane potential.
Therefore, if $g_s M^2 \gg N_{\overline{D3}}$ the flux potential is much larger than the antibrane contribution and the effect of the latter on the stabilization of $S$ can be ignored.
On the other hand, if $g_s M^2 \sim \mathcal{O}(N_{\overline{D3}})$ the effect of the antibrane can be potentially dangerous and can lead to a substantial shift in the vev of $S$.
This observation is compatible with the qualitative form of the bound \eqref{eq:Mbound}.

However, secondly, whether the shift in $S$ is large enough such that it can destabilize the conifold geometry also depends very much on the slope and qualitative shape of the two potentials.
Both are mostly determined by the behavior of the warp factor at the tip of the throat.
For the antibrane potential this is immediately apparent from \eqref{eq:VD3}.
The faster $e^{4A(0, S)}$ shrinks for $S\rightarrow 0$ the stronger the effect of the antibrane.
Importantly, since the integrand of \eqref{eq:explicitpotential} is approximatively localized in the IR region around small values of $\tau$ (see Figure~\ref{fig:integrandprofile}), the flux potential is also dominated by $e^{4A(0, S)}$.
If $e^{4A(0, S)}$ for $S\rightarrow 0$ tends towards zero fast enough (e.g.\ polynomial) the warp factor will overcome the fluxes and create another minimum at $S = 0$ as well as a local maximum somewhere in between.
In this case, if $g_s M^2$ is small enough, the combined potential \eqref{eq:combinedpotential} will not have a minimum at finite $S$ anymore but only at $S=0$ where $V=0$.
This is the situation that was assumed in the derivation of the potential \eqref{eq:VKS} with $e^{4A(0, S)} \sim S^{4/3}$, leading to the bound \eqref{eq:Mbound} on the existence of a runaway instability.

On the other hand, if $e^{4A(0, S)}$ does not go to zero for $S \rightarrow 0$ but stays finite, the flux potential goes to infinity and does not exhibit any further critical points.
Therefore, also the combined potential always has a minimum at $S>0$ and there in this analysis we see no runaway instability, independent on the value of $g_s M^2$.
This is the case for the potentials that we found in the previous section.

\begin{figure}[htb]
\centering
\begin{tabular}{cc}
\includegraphics[width=0.45\textwidth]{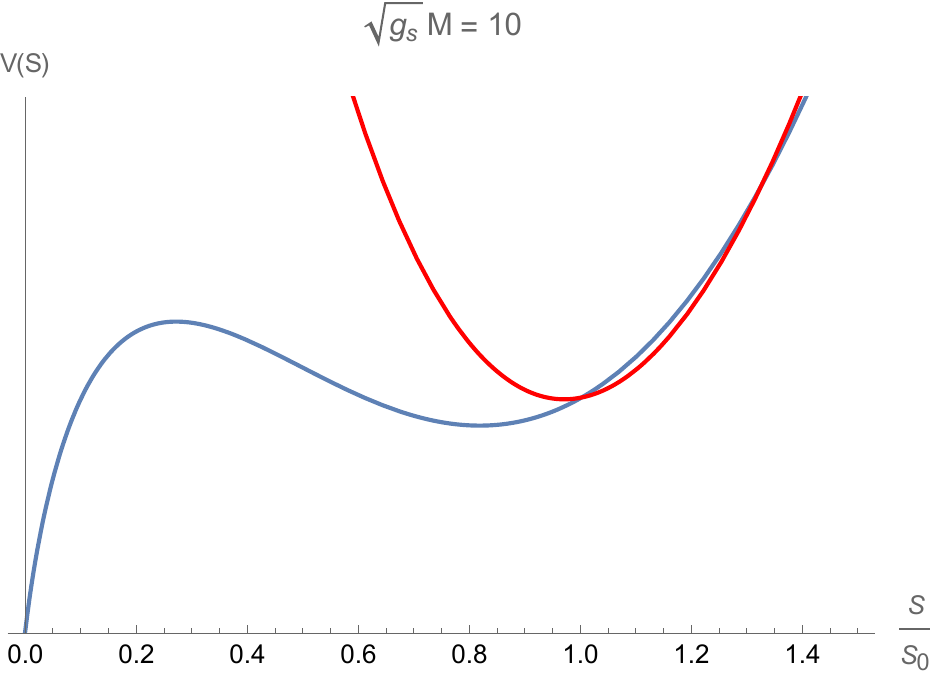}
\hspace{0.2em}&\hspace{0.2em}
\includegraphics[width=0.45\textwidth]{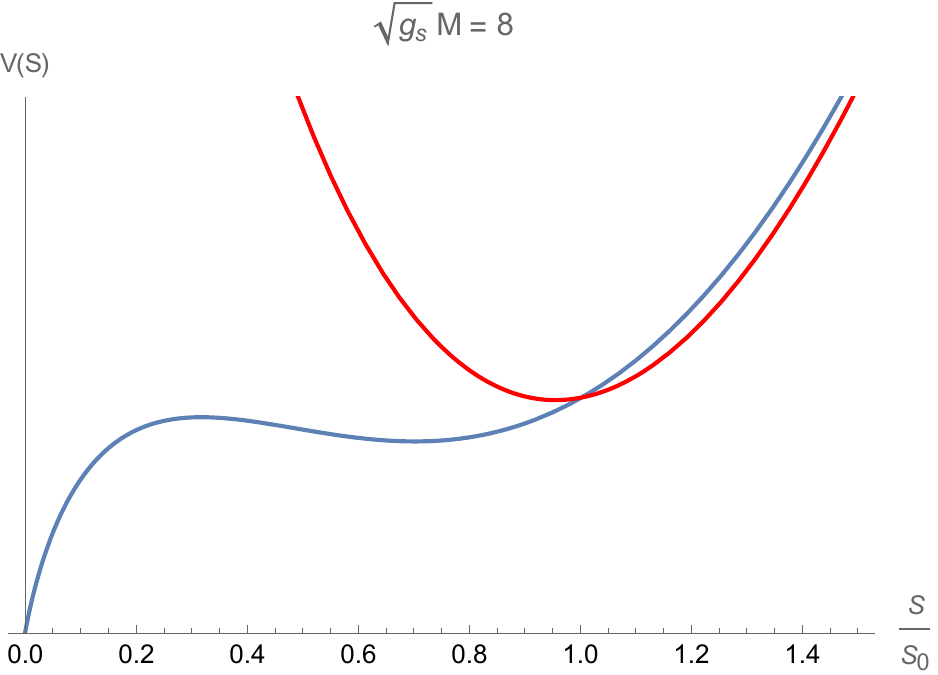}
\\[0.6em]
\includegraphics[width=0.45\textwidth]{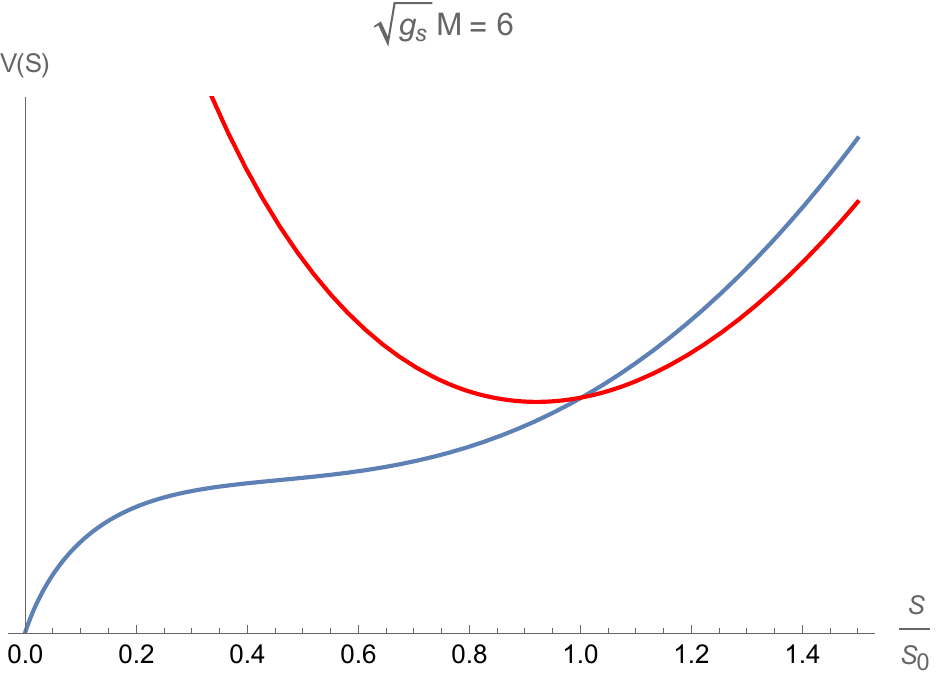}
\hspace{0.2em}&\hspace{0.2em}
\includegraphics[width=0.45\textwidth]{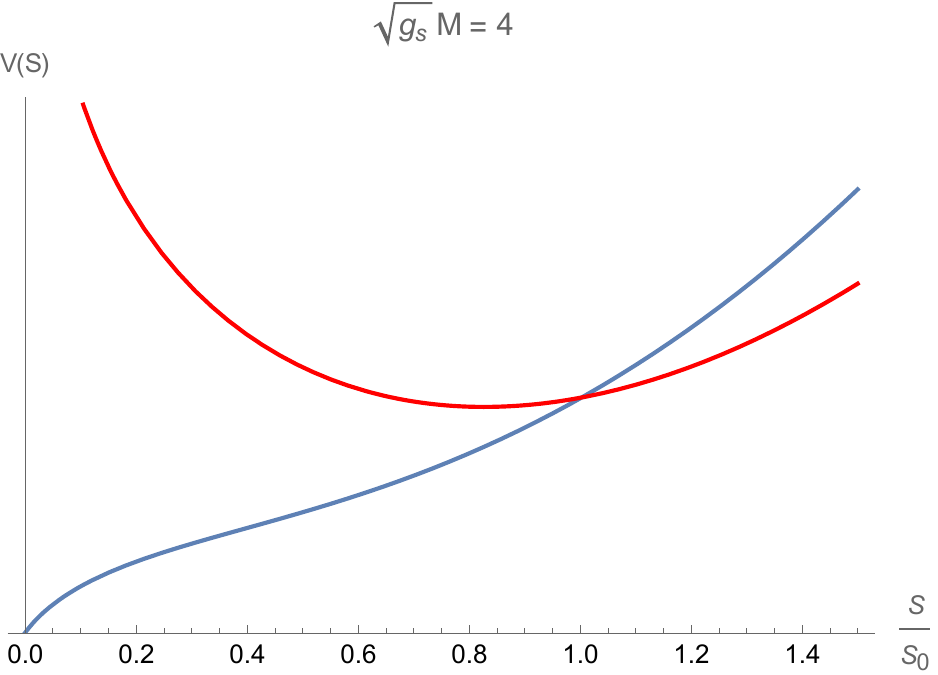}
\end{tabular}
\caption{Superposition of the potential for $S$ and the antibrane potential for different values of $g_s M^2$ (from large to small). The ``old'' potential is in blue, for small values of $M$ its minimum disappears. The red potential, which was computed by solving the constraint equations from Section~\ref{sec:warpeddef}, always has a minimum, irrespective of the value of $g_s M^2$. }
\label{fig:potentialsuperposition}
\end{figure}

The numerical result for the antibrane potential $V_{\overline{D3}}(S)$ together with the flux-potential $V_\mathrm{flux}$ for $S$ can be found in Figure~\ref{fig:potentialcomparison}.
This plot also contains the potentials that were used in \cite{Bena:2018fqc} (in blue) to argue that a $\overline{D3}$ brane can destabilize $S$ if the fluxes are too small.
We see that our flux-potential (in red) is not only more stable but also that our antibrane-potential is shallower and hence contributes less strongly to a potential destabilization for $S$.
We illustrate the superposition \eqref{eq:combinedpotential} of the two potentials in Figure~\ref{fig:potentialsuperposition} for different values of $g_s M^2$.
We see that even at values of $g_s M^2$ close to the critical value \eqref{eq:Mbound} the shift in the vev of $S$ is still relatively small.
Only if one goes to very small $g_s M^2$ would the effect of the antibranes be non-negligible.

\section{EFTs from warped compactifications}\label{sec:offshell}

The existence of the conifold runaway instability critically depends on the qualitative features of the flux potential $V_\mathrm{flux}$ in the region between $S=0$ and $S=S_0$.
The antibrane potential $V_{\overline{D3}}$ can only overrun $V_\mathrm{flux}$ and induce an instability if $V_\mathrm{flux}$ is bounded from above and does not go to infinity for $S\rightarrow 0$.
Therefore, knowing the potential only in the direct vicinity of $S_0$ is 
insufficient to check stability.
We have shown how to check the potential off-shell by including two constraint equations and shown the field S shifts only by a small amount with the addition of the antibrane. In this section we argue that a third constraint should be included in regions of $S$ farther from $S_0$.

Naively one would expect that one can compute the potential for any point in configuration space (consisting of the warp factor $A$, the internal metric $\tilde g_{mn}$ and other ten-dimensional fields) as the potential energy obtained by inserting the configuration into the ten-dimensional action and integrating over the compact dimensions.
This should not require solving any equations of motion as long as we are not interested in the actual dynamics of the system.
For example, a similar intuition would be correct for the classical mechanics of a single particle placed into a complicated potential.
One can assign a well defined potential energy to any point of the particles configuration space and would only have to solve its equations of motions if one wants to determine how it rolls down the potential.

However, as outlined in Section~\ref{sec:warpeddef}, in gravitational systems, due to their diffeomorphism invariance, the situation is substantially more complicated.
Some of the equations of motions have to be imposed as constraints and need to be satisfied along any consistent trajectory in configuration space.
A detailed account on how these constraints are related to the gauge freedom of general relativity can for example be found in \cite{Douglas:2008jx}.

In addition to the constraints we have already incorporated,  at finite distances from a minimum, i.e.~for deformations beyond linear order, another constraint has to be imposed, as was pointed out in \cite{Giddings:2005ff,Douglas:2009zn}.
After choosing a suitable three-dimensional slicing and time coordinate $t$ for the non-compact four-dimensional spacetime, the $G_{tt}$ component of the ten-dimensional Einstein equations does not contain any second derivatives $\partial_t^2$ and therefore cannot be satisfied dynamically.
This is the well-known Hamiltonian constraint of general relativity that has to be used to obtain a sensible lower-dimensional potential.

In particular, following \cite{Giddings:2005ff} we assume that there is a four-dimensional effective theory with space-time metric $\tilde g_{\mu\nu}$ and potential $V$.
We can consider an initial configuration where all velocities and second space derivatives vanish.
This is consistent as long as the second time derivatives are still unconstrained as they are determined by the equations of motion.
In this case the $tt$ component of the four-dimensional Einstein equations reduces to
\begin{equation}\label{eq:4deinsteinV}
    \tilde G_{tt} = \kappa_4^2 V \,,
\end{equation}
where $\tilde G_{tt}$ is the Einstein tensor of the four-dimensional metric $\tilde g_{\mu\nu}$ and $\kappa_4^2$ the four-dimensional gravitational constant.
This means that one can consistently determine $V$ by inserting \eqref{eq:4deinsteinV} back into the full ten-dimensional Einstein equations and solving for $V$.

In our case the relevant equation reads \cite{Giddings:2005ff}
\begin{equation}\label{eq:hamiltonianconstraint}
    \kappa_4^2 V = -\frac12 e^{4A} \tilde R_6 - \frac{1}{2} \tilde \nabla^2 e^{4A} + \frac14 e^{-4A} \left(\partial_m e^{4A} \partial^{\tilde m} e^{4A} + \partial_m \alpha \partial^{\tilde m} \alpha \right)   - e^{2A} \frac{1}{12 \Im \tau} G_{mnp} G^{\tilde m\tilde n\tilde p} \,.
\end{equation}
For the ansatz used in Section~\ref{sec:KSpotential} this expression can be simplified  further.
Here, by means of the constraint \eqref{eq:geometricconstraint2} the derivatives of $A$ and $\alpha$ are explicitly given in terms of \eqref{eq:integratedconstraint} and \eqref{eq:alphaoffshell} and $\tilde R_6 = 0$.
Inserting these relations into \eqref{eq:hamiltonianconstraint} yields
\begin{equation}\label{eq:GGconstraint}
    \tilde\star_6 \kappa_4^2 V =  \frac{e^{2A}}{12 \Im \tau} \Bigl[G_3 \wedge \tilde \star_6 G_3 - G_3 \wedge  G_3\Bigr] \,.
\end{equation}
After integrating over the internal dimensions and using that 
\begin{equation}
    \frac{1}{\kappa^2_4} = \frac{V_w}{\kappa^2_{10}}  = \frac{1}{\kappa^2_{10}} \int \tilde \star_6 e^{-4A} \,,
\end{equation}
we exactly reproduce our previous expression for the potential \eqref{eq:GGpotential}.
However, \eqref{eq:GGconstraint} poses also a constraint on the warp factor and internal metric because it must be satisfied even before integration.
This requirement is non-trivial since $\kappa_4^2 V$ is constant over the internal space while the right hand side of \eqref{eq:GGpotential} is a priori not.

Nonetheless, we argue that we can still trust our potential in the regime we are interested in even if it does not exactly satisfy \eqref{eq:GGconstraint}.
Firstly, it follows from the gravitational Bianchi identities that at first order in $S$ around the minimum the constraint \eqref{eq:hamiltonianconstraint} is actually equivalent to \eqref{eq:warpedoffdiagonal}.
This can be seen explicitly by taking the $S$ derivative of the right hand side of \eqref{eq:hamiltonianconstraint} and using the background equations of motion at $S=S_0$ to show that it vanishes identically as long as \eqref{eq:warpedoffdiagonal} holds.
The $S$ derivative of the left hand side of \eqref{eq:hamiltonianconstraint} must of course vanish at $S=S_0$ if $V$ is a sensible effective four-dimensional potential.
Notice also that $\partial_S \kappa^2_4 = 0$ as long as the trace-constraint \eqref{eq:warpedtraceless} is satisfied.
We therefore expect the potential that was computed in the previous sections to be a reasonably accurate approximation as long as it is evaluated close enough to its minimum.

On the other hand, this means that we 
do not know the form of the potential at small $S$ far from the supersymmetric minimum at at $S_0$, which in principle leaves open the possibility of a runaway instability at small $S$. 
This could be the case if $g_s M^2$ is very small, 
as then the effect of an antibrane would be large and it could generate a large displacement in $S$. 
However, in \cite{Bena:2018fqc} there was given an explicit critical value $M^2_\mathrm{min} \approx 46$ for $g_s M^2$ at which the addition of an antibrane would destabilize the Klebanov-Strassler geometry.
We can hence use our potential to compute the shift in $S$ away from the supersymmetric minimum at exactly this value and find
\begin{equation}
    \frac{S}{S_0} \approx 0.94 \,.
\end{equation}
This is still sufficiently close to one so we can rule out the existence of an instability of the previously described type at this value.
We illustrate this behavior in Figure~\ref{fig:potentialsuperposition}.

To get an accurate potential at very small values of $S$ and to really determine whether there can still be an instability at $g_s M^2 \ll M^2_\mathrm{min}$ we must modify our previous solutions such that they also satisfy \eqref{eq:GGconstraint}.
As for the potential \eqref{eq:GGpotential}, we obtain an explicit expression for the right hand side of \eqref{eq:GGconstraint} in terms of $\mathcal{T}(\tau, S)$ and $A(\tau, S)$ by using the Klebanov-Strassler solution for $G_3$.
The result will be essentially the same as the integrand of \eqref{eq:explicitpotential}, up to a different power of $e^A$ in the prefactor.
Solving this constraint and together with the two previously used constraints \eqref{eq:warpedtraceless} and \eqref{eq:warpedoffdiagonal}  is however not possible, since \eqref{eq:warpedtraceless} and \eqref{eq:warpedoffdiagonal} already determine the two functions $\mathcal{T}(\tau, S)$ and $A(\tau, S)$ uniquely.

It is hence necessary to extend the ansatz from Section~\ref{sec:KSpotential}  to solve all three constraints simultaneously.
There are in principle two different ways this can be done.
First, we can release the requirement \eqref{eq:constantfluxes} that the two-form fields $C_2$ and $B_2$ as well as the axio-dilaton $\tau_\mathrm{IIB}$ remain constant when varying $S$.
While integer quantization of the fluxes of course demands that the integrals of $F_3$ and $H_3$ over all three-cycles remain constant, this still retains the freedom to alter them by an exact piece, i.e.\ by changing the corresponding gauge fields $C_2$ and $B_2$.
In other words, $F_3$ and $H_3$ must remain constant in cohomology but they could be $S$-dependent as differential forms.
In our case the situation is slightly more subtle as we integrate $H_3$ over a non-compact cycle up to a UV-cutoff $\tau = \tau_\Lambda$.
Therefore changing $B_2$ can change $\int H_3$ by a boundary term, so we have to demand that $\delta B_2 = 0$ at $\tau = \tau_\Lambda$.

Allowing $B_2$ and $C_2$ to be $S$-dependent appears to be necessary from another point of view as well.
Namely, we have not commented yet on how to obtain a supersymmetric potential from \eqref{eq:GGconstraint}.
It is expected that in a IIB flux-compactification on a Calabi-Yau manifold the effective four-dimensional potential takes at the classical level the form of an $\mathcal{N} = 1$ no-scale potential
\begin{equation}\label{eq:noscale}
V = e^K g^{i\bar \jmath} D_i W \overline D_{\bar \jmath} \overline W \,,
\end{equation}
where $W$ is given by the GVW superpotential \cite{Gukov:1999ya}
\begin{equation}\label{eq:GVW}
    W = \int G_3 \wedge \Omega \,,
\end{equation}
with $\Omega$ the holomorphic three-form on the Calabi-Yau manifold.%
\footnote{See Appendix~\ref{app:cs} on how to construct $\Omega$ on the deformed conifold.}
Since $W$ is topological all non-trivial warping effects should be encoded in the K\"ahler potential $K$.
At trivial warp-factor the flux-potential \eqref{eq:GGconstraint} can be brought into the form \eqref{eq:noscale} by expanding $G^+_3 = G_3 + i \tilde \star_6 G_3$ in terms of a basis of harmonic forms on $H^3(\mathbb{C})$.\footnote{See e.g.~\cite{DeWolfe:2002nn} for a suggestion how to extend this procedure to the warped case.}
The covariant derivatives $D_i W$ then give the respective expansion coefficients.
This is of course  possible  only if $G^+_3$ is harmonic itself.
The harmonic representative in a given cohomology class however depends on the complex structure and hence differs for different values of $S$ by an exact piece.
This indicates again that $B_2$ and $C_2$ should be chosen $S$-dependent.

Another way to extend the family of backgrounds used in Section~\ref{sec:KSpotential} is to go away from the ansatz \eqref{eq:gmnofS} for the metric $\tilde g_{mn}$ and to allow it to become non-Ricci-flat.
This corresponds to exciting not only its complex-structure modulus (which is massless in the absence of fluxes) but also higher order Kaluza-Klein modes.
This could be plausible because their masses are also exponentially suppressed by the large warp factor and therefore in principle of the same order of magnitude as the mass of the complex-structure modulus.

Both approaches to extend the ansatz presumably need to be implemented simultaneously.
Moreover, they have the disadvantage that they destroy, at least individually, the  integrability of the constraint \eqref{eq:geometricconstraint2}.
We thus also do not expect \eqref{eq:integratedconstraint} and \eqref{eq:alphaoffshell} to hold anymore.
However, these relations were crucial in bringing the potential \eqref{eq:hamiltonianconstraint} into the familiar form \eqref{eq:GGconstraint}.
Therefore, it is conceivable that a potential for $S$ that satisfies all three constraints is not anymore of the pure flux form \eqref{eq:GGconstraint} but contains additional terms, e.g.~arising from KK-modes as also anticipated in \cite{Giddings:2005ff}.
It is however unclear how to bring such a potential in the form of $\mathcal{N} = 1$ supergravity with a classical superpotential given by \eqref{eq:GVW}, which remains an interesting subject for the future.

Nonetheless, the general lesson is clear. A consistent effective theory requires more than just studying the light zero modes when there is a warped compactification. The constraints can significantly affect the form of the metric including the warp factor in the infrared.

\section{Conclusions}\label{sec:conclusions}

In this paper we have studied the potential for light fields in the context of warped compactifications. Our results demonstrate that particular care is required for analyzing potentials and  stability.
Deriving the true lower-dimensional theory requires to account for the corrections induced by the warp factor to the flux-potential of all bulk complex structure moduli and to the K\"ahler moduli, including the volume modulus. 

In this paper we performed such a calculation by explicitly accounting for the constraint equations, which are nontrivial in every warped compactificiation.
In this way of performing the analysis, these constraints dictate the behavior of the warp factor when deforming the geometry of the compactification manifold away from the minimum of the potential.

Such a conclusion  might seem odd from a Wilsonian perspective in which we would expect we can analyze the low energy effective theory, ignoring what is assumed to be suppressed UV physics. And although we did not do the analysis from that perspective, we would nonetheless expect such a conclusion to be valid.  

The mass of the conifold modulus is indeed exponentially suppressed by the large hierarchy between the IR tip of the Klebanov-Strassler throat and the bulk Calabi-Yau in its UV.
This is nicely illustrated by the IR localized wave functions we computed numerically.
Consequently, the mass of the conifold modulus is hierarchically smaller than the masses of the remaining bulk complex structure moduli. However, it is comparable in mass to the light KK modes whose wavefunctions are concentrated in the infrared.

It is generally assumed that all complex structure moduli can be integrated out leaving an effective field theory for only the K\"ahler moduli in which their stabilization by non-perturbative effects can be studied.
However, in reality both the conifold deformation parameter as well as the IR-localized KK modes 
are exponentially light, so all such modes should be kept in a sensible effective description.

The reason the naive analysis is incorrect from this perspective is that the potential for $S$ neglected all the other light KK modes aside from the volume modulus, even though they are approximately the same in mass.

Therefore, an alternate way to interpret the modification of the warp factor in the infrared is that Kaluza Klein modes of the warped throat get turned on. This would change the volume of the compact directions from the naive warping behavior to the appropriate IR behavior consistent with the high-dimensional constraints. 

The reason this was not obvious in previous analyses is that the only light modes which were kept were the radion/conifold deformation parameter and the volume modulus (which is in fact parametrically lighter but naively more stable). Even if those are the only light zero modes, the warping means that all the moduli have light KK partners living at the tip of the conifold. A correct low energy analysis would have to include all of them.

This is a general lesson for warped compactifications, which occur fairly generically. The low energy degrees of freedom, which are presumably what will constitute the physics of the low energy world, will generally include many more modes than naively assumed.

 Our focus in this paper 
 was applying the formalism we have developed to the Klebanov-Strassler
 geometry and its interplay with the antibrane uplift in the KKLT scenario.  The conifold deformation parameter, which is  a complex structure modulus of the underlying Calabi-Yau geometry,  also plays the role of a radion in that it parameterizes the length of the throat and hence the hierarchy between its IR and UV scale and furthermore has the form of the potential expected for a radion \cite{Randall:2019ent}.
An antibrane placed in the IR of a strongly warped flux-background of this kind naturally tends to minimize its energy by increasing the IR-UV hierarchy.
It therefore drives the vacuum expectation value of the conifold deformation parameter to smaller values.
Whether this effect can eventually overcome the stabilization of the conifold by fluxes and create a runaway instability to a singular conifold depends both on the form and the magnitude of the flux-induced stabilization potential.

We have revisited the calculation of this potential and have found that the previously assumed behavior that was used to deduce the existence of an instability is too naive.
By numerically solving the relevant constraints we could explicitly determine the wave functions of the deformations of the internal metric and the warp factor and show that they are localized in the infrared. We furthermore  determined the off-shell dependence of the warp factor on the conifold deformation parameter 
and  found that this dependence is considerably weaker than previously assumed.
This implies that the effect of an antibrane on the stabilization of the conifold is also smaller than originally concluded.
However, to finally determine the fate of an antibrane in the Klebanov-Strassler geometry, we need to know the shape of the flux-potential at small values and hence finite displacements of the conifold modulus away from its supersymmetric minimum.
For large displacements, another constraint becomes relevant that we did not yet fully solve.
Nonetheless, we showed that with our potential, even at flux values where previous works expected the onset of a runaway instability, the induced shift in conifold deformation parameter is relatively small.
Therefore it seems to be unlikely that higher order effects that are needed to solve the last constraint drastically change the potential in the relevant regimes.
Our results therefore indicate that the addition of an antibrane in the IR of the Klebanov-Strassler geometry does not destabilize the underlying conifold geometry or at worst only at very small fluxes.

Interestingly, solving  the remaining constraint exactly seems to require to modify our ansatz in a fundamental way.
Such modifications can include a non-trivial dependence of the IIB two-form gauge fields on the conifold deformation parameter, a non-constant axio-dilaton, as well as deformations of the internal metric that drive it away from Ricci-flatness.
While the former indicates a non-trivial coupling to the K\"ahler moduli sector (see also \cite{Martucci:2016pzt}), 
the latter two correspond to excitations of higher Kaluza-Klein modes of the Calabi-Yau metric and the IIB axio-dilaton.
We intend to revisit these issues in future work.

\section*{Acknowledgements}
We would like to thank Iosif Bena, Ralph Blumenhagen, Emilian Dudas, Hao Geng, Mariana Gra\~na, Arthur Hebecker, Luca Martucci, Rashmish Mishra, Jakob Moritz, and Thomas van Riet for very helpful discussions and correspondence.
We would like to thank Andrew Frey and Ivonne Zavala for pointing out a typo in a previous version of the paper.
The work of SL is supported by the NSF grant PHY-1915071.
The work of LR is supported by NSF grants PHY-1620806 and PHY-1915071, the Chau Foundation HS Chau postdoc award, the Kavli Foundation grant “Kavli Dream Team,” and the Moore Foundation Award 8342.

\newpage

%%%%%%%%%%%%%%%%%%%%%%%%%%%%%%%%%%%%%%%%%%%%%%%%%%%%%%%%%%%

%\newpage

\appendix
\noindent
{\bf\Huge Appendix}

\section{Frames on the deformed conifold}\label{app:conifoldframes}

 The (deformed) conifold has an $SU(2) \times SU(2)$ isometry.
 To find a metric we introduce the Euler-angles $(\psi_1, \phi_1, \theta_1)$ and $(\psi_2, \phi_2, \theta_2)$ on the two $SU(2)$s and the one-forms \cite{Papadopoulos:2000gj}
 \begin{equation}\begin{gathered}
e^1 = \sin \frac{\psi}{2} \sin \theta_1 d {\phi_1} + \cos \frac{\psi}{2} d \theta_1 \,,\qquad
e^2 = - \cos \frac{\psi}{2} \sin \theta_1 d {\phi_1} + \sin \frac{\psi}{2} d \theta_1 \,, \\
\tilde e^1 = \sin \frac{\psi}{2} \sin \theta_2 d {\phi_2} + \cos \frac{\psi}{2} d \theta_2 \,,\qquad
\tilde e^2 = - \cos \frac{\psi}{2} \sin \theta_2 d {\phi_2} + \sin \frac{\psi}{2} d \theta_2 \,, \\[0.4em]
e^3 + \tilde e^3 = d \psi + \cos \theta_1 d\phi_1 + \cos \theta_2 d\phi_2 \,,
\end{gathered}\end{equation}
where $\psi = \psi_1 + \psi_2$.
They are left invariant under the respective $SU(2)$s and satisfy the Maurer-Cartan equations $d e^i = \frac 12 \epsilon^i{}_{jk} e^j \wedge e^k$ and $d \tilde e^i = \frac 12 \epsilon^i{}_{jk} \tilde e^j \wedge \tilde e^k$.
After introducing the following linear combinations
\begin{equation}\begin{gathered}\label{eq:gframe}
g^1 = \frac{1}{\sqrt{2}} \left(e^1 - \tilde e^1 \right) \,,\qquad
g^2 = \frac{1}{\sqrt{2}} \left(-e^2 - \tilde e^2 \right) \,,\qquad
g^3 = \frac{1}{\sqrt{2}} \left(e^1 + \tilde e^1 \right) \,, \\
g^4 = \frac{1}{\sqrt{2}} \left(-e^2 + \tilde e^2 \right) \,,\qquad
g^5 = e^3 + \tilde e^3 \,.
\end{gathered}\end{equation}
In this frame the metric on the singular conifold reads
\begin{equation}
ds^2 = d r^2 + r^2 ds_{T^{1,1}}^2 \,, 
\end{equation}
where
\begin{equation}\label{eq:T11metric}
 ds_{T^{1,1}}^2 = \frac16 \Bigl[(g^1)^2 +(g^2)^2 + (g^3)^2 + (g^4)^2\Bigr] + \frac19 (g^5)^2 \,,
\end{equation}
and $r$ is a sixth, radial coordinate. 

When reducing the ten-dimensional IIB action to four dimensions one also has to account for the volume of the five-dimensional transverse space.
The according factor can be obtained by direct integration over the domain $0 \leq \theta_i \leq \pi$, $0 \leq \phi_i \leq 2 \pi$ and $0 \leq \psi \leq 4\pi$ and is given by
\begin{equation}\label{eq:5dvolume}
    \mathrm{vol}_5 = \int g^1 \wedge g^2 \wedge g^3 \wedge g^4 \wedge g^5 = 64 \pi^3 \,.
\end{equation}
This differs from the volume of the volume of the transverse space of the singular conifold, namely $T^{1,1}$ with the metric \eqref{eq:T11metric} by a numerical factor,
\begin{equation}
    \mathrm{vol}_{T^{1,1}} = \frac{\mathrm{vol}_5}{108} = \frac{16}{27} \pi^3 \,.
\end{equation}

\section{Complex structure deformations of the deformed conifold}\label{app:cs}

In this appendix we discuss how the deformation parameter $S$ is related to a complex structure deformation of the  Calabi-Yau structure of the deformed conifold.
In particular, in the main text we have treated $S$ as a real variable, even though it is in fact complex.
Here, we illustrate that the related variable from the main text is actually the absolute value $|S|$ of the complex conifold deformation parameter and how to correctly describe variations with respect to $S$ and its complex conjugate $\bar S$.

For this purpose we first need to define a suitable complex basis on the cotangent space.
Following \cite{Papadopoulos:2000gj} we introduce the following set of forms,
\begin{equation}\begin{aligned}
E^1 &= \frac{1}{\sqrt{2}} e^{\frac{x(\tau)}{2}} \left(e^{\frac{y(\tau)}{2}} g^1 - e^{-\frac{y(\tau)}{2}} g^3 \right) \,, \qquad
E^2 = \frac{1}{\sqrt{2}} e^{\frac{x(\tau)}{2}} \left(-e^{\frac{y(\tau)}{2}} g^2 - e^{-\frac{y(\tau)}{2}} g^4 \right) \,, \\
E^3 &= \frac{1}{\sqrt{2}} e^{\frac{x(\tau)}{2}} \left(-e^{\frac{y(\tau)}{2}} g^1 + e^{-\frac{y(\tau)}{2}} g^3 \right) \,, \qquad
E^4 = \frac{1}{\sqrt{2}} e^{\frac{x(\tau)}{2}} \left(-e^{\frac{y(\tau)}{2}} g^2 + e^{-\frac{y(\tau)}{2}} g^4 \right) \,, \\
E^5 &= e^{-3 p(\tau) - \frac{x(\tau)}{2}} d \tau \,, \qquad
E^6 = e^{-3 p(\tau) - \frac{x(\tau)}{2}} g^5 \,, \\
\end{aligned}\end{equation}
where $p$, $x$ and $y$ are the same functions as in the metric \eqref{eq:conifoldmetricansatz} which in this frame becomes $ds^2 = \delta_{ab} E^a E^b$.
Moreover, we introduce the complex frame
\begin{equation}\label{eq:complexframe}
Z^1 = E^1 + I E^2 \,,\qquad Z^2 = E^3 + I E^4 \,,\qquad Z^3 = E^5 + I E^6 \,,
\end{equation}
together with their complex conjugates $\bar Z^{\bar 1, \bar 2, \bar 3}$.
In this frame the K\"ahler form of the deformed conifold is given by
\begin{equation}
\omega = \frac i2 \left( Z^1 \wedge \bar Z^{\bar 1} + Z^2 \wedge \bar Z^{\bar 2}+ Z^3 \wedge \bar Z^{\bar 3} \right) \,,
\end{equation}
and the holomorphic three-form reads
\begin{equation}
\Omega = Z^1 \wedge Z^2 \wedge Z^3 \,.
\end{equation}
With respect to the basis \eqref{eq:gframe} they become
\begin{equation}\begin{aligned}\label{eq:CYformsansatz}
\omega &= e^{-6p-x} d\tau \wedge g^5 - e^x \left(g^1 \wedge g^4 - g^2 \wedge g^3 \right)  \,, \\
\Omega &= e^{-3p-\frac{x}{2}} \left(d \tau + i g^5\right) \wedge \left(g^1 \wedge g^3 + g^2 \wedge g^4 - i e^y g^1 \wedge g^2 + i e^{-y} g^3 \wedge g^4 \right) \,.
\end{aligned}\end{equation}
The Calabi-Yau conditions
\begin{equation}
    d \omega = d \Omega = 0 \,,
\end{equation}
translate into a system of first-order differential equations for $p$, $x$ and $y$,
\begin{equation}\begin{aligned}
\dot x  &= e^{-6p -2 x} \,, \\
6 \dot p- \dot x  &= 2 \cosh y \,, \\
\dot y &= - \cosh y \,,
\end{aligned}\end{equation}
where the dot denotes differentiation with respect to $\tau$.
This is the same set of first order equations \eqref{eq:susyeqs} as resulting from the one-dimensional Lagrangian \eqref{eq:1dlagrangian} and is readily solved by \eqref{eq:conifoldsol}.
This shows that the metric \eqref{eq:defconifoldmetric} is indeed Calabi-Yau.
Inserting the solution \eqref{eq:conifoldsol} for $p$, $x$ and $y$ (with $|c_1| = \left|S\right|^{2/3}$ and $c_2 = c_3 =0$) into \eqref{eq:CYformsansatz} finally gives
\begin{equation}\begin{gathered}
    \omega = \left|S\right|^2\frac{K(\tau)}{2} \left[ \frac{1}{3 K(\tau)^3} d\tau \wedge g^5 - \frac{\sinh(\tau)}{2} \left(g^1 \wedge g^4 - g^2 \wedge g^3 \right)\right] \,, \\
    \Omega = S \, \frac{\sinh \tau}{4\sqrt{6}}  \left(d \tau + i g^5\right) \wedge \left(g^1 \wedge g^3 + g^2 \wedge g^4 - i \tanh \frac\tau2 \, g^1 \wedge g^2 + i \coth \frac\tau2 \, g^3 \wedge g^4 \right) \,.
\end{gathered}\end{equation}
Notice, that here we treat $S$ as complex and that $\omega$ depends only on its absolute value while $\Omega$ depends holomorphically on it.
$\bar \Omega$ can be obtained from $\Omega$ by complex conjugation.

We now want to determine how $\omega$ and $\Omega$ behave under a variation of $S$.
In Section~\ref{sec:unwarped} we argued that such a variation has to be accompanied by a suitable diffeomorphism.
We defined the gauge-fixed variation with respect to $\left|S\right|$ as
\begin{equation}
\delta_{\left|S\right|} = \partial_{\left|S\right|} - \frac{2}{ \left|S\right|} \mathcal{L}_\eta \,,
\end{equation}
where $\eta$ is the vector field
\begin{equation}
    \eta = \frac{\sinh 2\tau - 2\tau}{4 \sinh^2 \tau} \partial_\tau \,.
\end{equation}
We now want to extend this to a holomorphic variation.
Therefore we use the complex structure on the deformed conifold to introduce another vector field,
\begin{equation}
    \xi = \iota_\eta \omega = \frac{\sinh 2\tau - 2\tau}{4 \sinh^2 \tau} \partial_\psi \,.
\end{equation}
In the limit $\tau \rightarrow \infty$ where the metric of the deformed conifold becomes that of a cone over the Sasaki-Einstein manifold $T^{1,1}$ the first vector field goes to the homothetic vector field $\eta \rightarrow \partial_\tau \sim r \partial_r$ and $\xi$ is called the Reeb or characteristic vector field on $T^{1,1}$ (see e.g.~\cite{Sparks:2010sn}).
The latter is a Killing vector field of $T^{1,1}$ from which it follows that $\mathcal{L}_\xi g \rightarrow 0$.

Using these two vector fields we can introduce the holomorphic and anti-holomorphic variations
\begin{equation}\label{eq:csdeformation}
\delta_S = \partial_S - \frac1S \mathcal{L}_{\eta - i \xi} \,,\quad\quad 
\delta_{\bar S} = \partial_{\bar S} - \frac{1}{\bar S} \mathcal{L}_{\eta + i \xi} \,.
\end{equation}
To show that this is a sensible definition we compute the action of $\delta_S$ and $\delta_{\bar S}$ on $\omega$ and $\Omega$.
A direct computation reveals that both act trivially on $\omega$,
\begin{equation}
    \delta_S \omega = \delta_{\bar S} \omega = 0 \,,
\end{equation}
hence $\delta_S$ is not a K\"ahler transformation.
Moreover, using Cartan's formula one immediately finds that
\begin{equation}
\delta_{\bar S} \Omega = d \left(\iota_{\eta- i \xi} \Omega \right) = 0 \,,
\end{equation}
and analogously $\delta_S \overline \Omega = 0$.

A bit more work is required to determine the action of $\delta_S$ on $\Omega$.
The result is given by
\begin{equation}\label{eq:chiS}
    \chi_S \equiv \delta_S \Omega = \frac{1}{2 \sqrt{6}} \Bigl( g^3 \wedge g^4 \wedge g^5 + d \left[ F(\tau) \left(g^1 \wedge g^3 + g^2 \wedge g^4 \right)  \right] - i d \left[f(\tau) g^1 \wedge g^2 + k(\tau) g^3 \wedge g^4 \right] \Bigr) \,,
\end{equation}
where the functions $F$, $f$ and $k$ are given by the Klebanov-Strassler solution,
\begin{equation}\begin{aligned}
    F(\tau) &= \frac{\sinh \tau - \tau}{2 \sinh \tau} \,,\\
    f(\tau) &= \frac{\tau \coth \tau - 1}{2 \sinh \tau} \left(\cosh \tau -1\right) \,,\\
    k(\tau) &= \frac{\tau \coth \tau - 1}{2 \sinh \tau} \left(\cosh \tau +1\right) \,.
\end{aligned}\end{equation}
With respect to the complex basis \eqref{eq:complexframe} $\chi_S$ reads
\begin{equation}
\chi_S = \frac{1}{\sinh^2 \tau} \left [(\tau \coth \tau -1) Z^1 \wedge Z^2 \wedge \bar Z^{\bar3} + \frac{\sinh 2\tau - 2\tau}{4 \sinh \tau} \left(Z^1 \wedge Z^3 \wedge \bar Z^{\bar1} - Z^2 \wedge Z^3 \wedge \bar Z^{\bar2}\right)\right] \,,
\end{equation}
and is therefore indeed a $(2,1)$ form. This confirms that \eqref{eq:csdeformation} acts as a complex structure deformation on $\Omega$.
Moreover, a direct computation shows that
\begin{equation}
    \delta_S g_{mn} = \tfrac18\bar \Omega_{m}{}^{kl} \left(\chi_S\right)_{nkl} \,.
\end{equation}

  We can also use these expression to compute the periods of $\Omega$.
  The A-cycle period is most easily evaluated at $\tau = 0$ where the A-cycle, corresponding to the $S^3$ at the tip of the conifold, is spanned by $g^3 \wedge g^4 \wedge g^5$.
  Therefore, it is given by
  \begin{equation}
       \int_A \Omega \sim S \,,
  \end{equation}
  where we suppressed a constant normalization factor.
 Meanwhile, the B-cycle is non-compact and spanned by $\frac12 d \tau \wedge (g^1 \wedge g^2 + g^3 \wedge g^4)$.
 Using the regulator \eqref{eq:Lambda0} and suppressing the same constant factor gives the period integral
  \begin{equation}
      \int_B \Omega \sim  \frac{1}{2\pi i} S \log \frac{\Lambda_0^3}{S} + \mathcal{O(S)} \,.
  \end{equation}
In a similar fashion one can obtain the periods of $\chi$ from \eqref{eq:chiS}.
  
In Section~\ref{sec:unwarped} we were able to integrate $\eta$ and describe the implicit $\left|S\right|$-dependence that it encodes in terms of an explicitly $\left|S\right|$-dependent $\tau$-coordinate, see \eqref{eq:finitetauchange}.
However, it is not possible to do the same for $S$ and $\bar S$ simultaneously.
This follows from the fact that $\xi$ and $\eta$ do not commute, and hence the distribution spanned by $\xi + i \eta$ and $\xi - i \eta$ is non-integrable.
Therefore, it is not possible to find a set of new coordinates
\begin{equation}\begin{aligned}
    \tau &\rightarrow \mathcal{T}(\tau, S, \bar S) \,, \\
    \psi &\rightarrow \Psi(\tau, S, \bar S) \,,
\end{aligned}\end{equation}
that are functions of both $S$ and $\bar S$ and such that the action of $\delta_S$ and $\delta_{\bar S}$ on the metric as well as $\omega$ and $\Omega$ reduces to simple derivatives $\partial_S$ and $\partial_{\bar S}$.

\section{Comparison with the hard-wall approximation}\label{app:hardwall}

In Section~\ref{sec:KSpotential} we chose an ansatz for the deformed metric $\tilde g_{mn}$ in \eqref{eq:gmnofS} that preserves the diagonal structure of \eqref{eq:defconifoldmetric}.
This is done by replacing the radial coordinate $\tau$ in the metric on the deformed conifold by an $S$ dependent coordinate $\mathcal{T}(\tau, S)$.
Infinitesimally, this corresponds to including a gauge compensator $\eta^m$ in the deformation of the metric, $\delta_S \tilde g_{mn} = \partial_S \tilde g_{mn} + (\mathcal{L}_\eta g)_{mn}$, such that $\eta^m$ has only a non-vanishing component in the $\tau$-direction, cf.~\eqref{eq:etaansatz}.
However, as we have seen in the previous appendix for the unwarped case, it makes in principle sense to also include a gauge compensator in the $\psi$-direction.
In this appendix we extend this discussion to the case of a non-trivial warp-factor $A$ and argue that including $\eta^\psi$ is not relevant for our previous analysis.
Moreover, we compare this ansatz with the computation in the  hard-wall approximation in \cite{Douglas:2008jx}.

We consider infinitesimal deformations of the metric of the deformed conifold of the form
\begin{equation}
    \delta_S \tilde g_{mn} = \partial_S \tilde g_{mn} + \nabla_m \eta_n + \nabla_n \eta_m \,,
\end{equation}
where $\tilde g_{mn}$ is given by \eqref{eq:defconifoldmetric} and for the compensator $\eta^m$ we take the ansatz
\begin{equation}
    \eta^m = \bigl( \eta^\tau(\tau), 0, 0,0,0, \eta^\psi(\tau) \bigr ) \,,
\end{equation}
with resepect to the dual basis of \eqref{eq:gframe}.
Importantly, $\eta^\psi(\tau)$ drops out of the first constraint \eqref{eq:warpedtraceless} which yields the following first order equation for $\eta^\tau(\tau)$
\begin{equation}\label{eq:etataueqlin}
    \frac{1}{S} + \coth \tau \, \eta^\tau  +2 \partial_\tau \eta^\tau - 2 \delta_S A = 0 \,.
\end{equation}
Moreover, in the second constraint, most conveniently analyzed in its form \eqref{eq:warpedharmonic}, $\eta^\tau(\tau)$ and $\eta^\psi(\tau)$ decouple, resulting in two independent second order equations,
\begin{multline}\label{eq:etataueqquad}
\partial_\tau^2 \eta^\tau + 2 \coth \tau \, \partial_\tau \eta^\tau -2 \operatorname{csch}^2 \tau \, \eta^\tau \\
- \frac43 \partial_\tau A \left(\frac{2}{S} + 6 \partial_\tau \eta^\tau + \frac{9 + \operatorname{csch}\tau \sinh 3\tau - 12 \tau \coth \tau}{\sinh 2 \tau - 2 \tau} \eta^\tau\right)  = 0\,, 
\end{multline}
and 
\begin{equation}\label{eq:etapsieq}
\partial_\tau^2 \eta^\psi + 2 \coth \tau \, \partial_\tau \eta^\psi -2 \operatorname{csch}^2 \tau \, \eta^\psi - 4 \partial_\tau A \, \eta^\psi = 0 \,.
\end{equation}
This means that we can solve for $\eta^\tau(\tau)$ and $\eta^\psi(\tau)$ completely independently from each other.
Moreover, $\delta_S A$ does not enter the equation \eqref{eq:etapsieq} for $\eta^\psi(\tau)$, hence the presence of a non-trivial $\eta^\psi(\tau)$ does neither affect $\eta^\tau(\tau)$ nor the $S$-dependence of the warp factor $A$.
According to the results of Appendix~\ref{app:cs} this means that $A$ depends only on the absolute value of $S$ but not on its phase.
The equations for $\eta^\tau(\tau)$, \eqref{eq:etataueqlin} and \eqref{eq:etataueqquad}, can be obtained from the integrated constraints \eqref{eq:volumeconstraintsol} and \eqref{eq:harmonicconstraintsol} by taking their derivative with respect to $S$ and identifying $\eta^\tau$ with $\partial_S \mathcal{T}$.
Therefore they yield--at least infinitesimally--the same solution as we have constructed numerically in Section~\ref{sec:KSpotential}. 

In \cite{Douglas:2008jx} the equations \eqref{eq:etataueqlin} to \eqref{eq:etapsieq} where solved in the so-called hard-wall approximation.
There, the relatively complex Klebanov-Strassler solution is replaced by a simple AdS background and its smooth cap-off in the IR is mimicked by a hard cut-off of the radial coordinate.
This means the internal metric is the one of the singular conifold (i.e.~a cone over $T^{1,1}$),
\begin{equation}
ds_6^2 = \frac{3}{2^{5/3}} S^{2/3} e^{2 \tau /3} \left[ \frac19 \left(d \tau^2 + (g^5)^2 \right) + \frac16 \sum_{i = 1}^4 (g^i)^2  \right]  \,,
\end{equation}
and $A$ takes the form of a pure AdS warp factor,
\begin{equation}\label{eq:AdSwarpfactor}
    e^{-4A} \propto \frac{g_s (\alpha' M)^2}{\left|S\right|^{\frac43}} e^{-4\tau/3} \,.
\end{equation}
Notice, that this expression differs from its asymptotic value in the KS solution (described by the Klebanov-Tseytlin solution \cite{Klebanov:2000nc}) by a missing factor of $\left(\tau - \frac14\right)$.
In this case the equations for $\eta^\tau(\tau)$ reduce to
\begin{equation}\begin{aligned}
    \frac{4}{ S} + \frac{4}{3} \eta^\tau (\tau) + 2 \partial_\tau \eta^\tau &= \partial_S A  \,,\\
    \frac{8}{9 S} + \frac{8}{9 } \eta^\tau + \frac{2}{3} \partial_\tau \eta^\tau (\tau) - \partial_\tau^2 \eta^\tau &= 0 \,,
\end{aligned}\end{equation}
where we used $\delta_S A = \partial_S A + \partial_\tau A \eta^\tau$ and the equation for $\eta^\psi(\tau)$ becomes
\begin{equation}
\frac23 \partial_\tau \eta^\tau + \partial_\tau^2 \eta^\tau = 0 \,.
\end{equation}
These equations are solved by
\begin{equation}\label{eq:hardwalletatau}
    \eta^\tau(\tau) = - \frac{1}{S} + a_1 e^{-2\tau/3}  \,,\qquad  
    \eta^\psi(\tau) =  b_1 + b_2  e^{-2\tau/3} \,,
\end{equation}
with integration constants $a_1$, $b_1$ and $b_2$ and
\begin{equation}\label{eq:hardwalldA}
    \partial_S A = \frac{1}{3 S} \,.
\end{equation}
This is the solution found in \cite{Douglas:2008jx}.
Note, in particular, that \eqref{eq:hardwalldA} is compatible with the $e^{-4A} \propto S^{-4/3}$ scaling of the warp factor (however $\delta_S A = 0$).
Moreover, in this case the two equations in \eqref{eq:hardwalletatau} are equivalent, the second equation can be obtained by taking the derivative of the first one.
This structure, however, disappears as soon as the warp factor deviates from the pure AdS form \eqref{eq:AdSwarpfactor}, for example by reinstating the factor of $\left(\tau - \frac14\right)$ found in the Klebanov-Tseytlin solution or for the full Klebanov-Strassler warp factor.
Hence, the $S^{-4/3}$ scaling found in \cite{Douglas:2008jx} seems to be indeed an accident of the hard-wall approximation.

%%%%%%%%%%%%%%%%%%%%%%%%%%%%%%%%%%%%%%%%%%%%%%%%%%%%%%%%%%%
%\newpage

\providecommand{\href}[2]{#2}\begingroup\raggedright\endgroup

\end{document}